\documentclass[11pt]{article}

\usepackage{amsmath}
\usepackage{graphicx}
\usepackage{indentfirst}
\usepackage{amssymb}
\usepackage{cite}
\usepackage{color}
\usepackage{subfigure}
\usepackage{xcolor}
\usepackage[breaklinks=true]{hyperref}

\begin{document}

\title{  Shadow thermodynamics of the Hayward-AdS black hole}

\date{}
\maketitle

\begin{center}
	Zhi Luo$~^{a,}$\footnote{ZhiLuo@cqu.edu.cn},~Hao Yu$~^{a,b,}$\footnote{yuhaocd@cqu.edu.cn},~Shuo Cao$~^{c,}$\footnote{caoshuo@bnu.edu.cn},~Jin Li~$^{a,b,}$\footnote{cqstarv@hotmail.com, corresponding author} \\
\end{center}

\begin{center}
	$^a$ Physics Department, Chongqing University, Chongqing 401331, China\\
	$^b$ Department of Physics and Chongqing Key Laboratory for Strongly Coupled Physics, Chongqing University, Chongqing 401331, China\\
	$^c$ Department of Astronomy, Beijing Normal University, Beijing 100875, China
\end{center}

\vskip 0.6in
{\abstract
{
{In this paper, the phase structure of the Hayward-AdS black hole (BH) is studied using shadow formalism. It has been found that the shadow radius is a monotonic function of the horizon radius and can therefore play an equivalent role to the horizon radius in characterizing the thermodynamics of Hayward-AdS BH. The thermodynamic phase transition (PT) of the Hayward-AdS BH is investigated with the shadow radius. It is shown that as the magnetic charge increases, the shadow radius becomes larger, while the coexistence temperature becomes lower. The thermal profile of the Hayward-AdS BH is established by combining the temperature diagram and the shadow cast diagram, which shows that for a fixed magnetic charge, the temperature of the Hayward-AdS BH increases with the pressure while the region of the thermal profile decreases with the pressure. In particular, the temperature of the Hayward-AdS BH follows an N-type change trend when it is smaller than the critical temperature. This imply that the BH shadow may be used to investigate the thermodynamics of the Hayward-AdS BH.}
}

%%%

\thispagestyle{empty}
\newpage
\setcounter{page}{1}

\section{Introduction}\label{sec1}
BHs are special celestial bodies with strong gravitational fields. Since the general relativity has been proposed, the theoretical research and observations on BHs have been widely concerned. In 2015, the Laser Interferometer Gravitational-Wave Observatory (LIGO) observed the first gravitational wave event generated by the merger of binary BHs, which provides strong evidence for the existence of BHs~\cite{LIGOScientific:2016sjg,LIGOScientific:2016vlm,LIGOScientific:2016aoc}. On the other hand, it is widely believed that there exist supermassive BHs at the centers of most galaxies. Therefore, the shadow images of the supermassive BHs at the centers of the supergiant elliptical galaxy $M87^{\star}$ and the Milky Way, released by Event Horizon Telescope (EHT), can be regarded as direct evidence for the existence of BHs and the validity of the general relativity \cite{EventHorizonTelescope:2019dse,EventHorizonTelescope:2019uob,EventHorizonTelescope:2019jan,EventHorizonTelescope:2019ths,EventHorizonTelescope:2019pgp,EventHorizonTelescope:2019ggy,EventHorizonTelescope:2022xnr,EventHorizonTelescope:2022vjs,EventHorizonTelescope:2022wok,EventHorizonTelescope:2022exc,EventHorizonTelescope:2022urf,EventHorizonTelescope:2022xqj}. The shadow images reveal primordial information on the dynamics of the jets and matter around the BHs, and thus with the shadow images one can constrain the physical parameters of the BHs, such as mass, rotation and charge. Moreover, the shadow images are also affected by the theory of gravity, so they can be used to test the validity of various modified theories of gravity.

There is an intense gravitational effect in the vicinity of a BH, which causes nearby objects to fall into the BH when they are within the critical radius. Thus, not all the light passing through a BH can be received by the distant observer. The photons with  particular incidence angles will be attracted into the BH, and then generate the BH shadow (a two-dimensional dark zone in the celestial sphere). The bright ring around the BH shadow is composed of the photons bent by the gravitational field, which is called the photon ring. For different BHs, the corresponding shadow images have different properties. The shadow of the Schwarzschild BH is the simplest case, which is related to the BH mass and the location of the observer~\cite{Synge:1966okc,Luminet:1979nyg}. Due to the dragging effect, the shadow of the rotating Kerr BH is deformed~\cite{Vries1,Hioki:2008zw,Eiroa:2017uuq}. For the past few years, the shadows of various BHs have been investigated~\cite{Bambi:2008jg,Bambi:2010hf,Atamurotov:2013sca,Papnoi:2014aaa,Atamurotov:2015nra,Wang:2017qhh,Guo:2018kis,Yan:2019etp,Konoplya:2019sns}. For an exhaustive overview of the BH shadow, one can refer to Ref.~\cite{Cunha:2018acu}. 

Not only does the BH shadow provide evidence for the existence of BHs, it also can be used to test the existence of the thermodynamic phase transition (PT) of the AdS BH. In the 1970s, Hawking and Bekenstein et al., pointed out that a BH can be regarded as a thermodynamic system because they have numerous similarities~\cite{Hawking:1975vcx,Hawking:1976de,Bekenstein:1974ax}. Before long, the PT between the Schwarzschild AdS BH and the thermal AdS space was first found by Hawking and Page~\cite{Hawking:1982dh}. In recent years, the thermodynamics of the AdS BH has been a widely topic in physics due to the AdS/CFT correspondence~\cite{Maldacena:1997re,Witten:1998qj,Gubser:1998bc,Cvetic:2010jb,Chamblin:1999tk,Chamblin:1999hg,Caldarelli:1999xj}. For an AdS BH, it has been found that the negative cosmological constant can be seen as the thermodynamic pressure in the extended phase space~\cite{Kastor:2009wy,Dolan:2010ha}. The van der Waals (vdW)-like PT of the charged AdS BH was studied in Ref.~\cite{Kubiznak:2012wp}, which shares a resemblance with the liquid-gas PT. More research on the PT (for example, the reentrant PT and the triple point PT) of the AdS BH can be found in Refs.~\cite{Gunasekaran:2012dq,Altamirano:2013uqa,Altamirano:2013ane,Frassino:2014pha,Wei:2014hba,Hendi:2014kha,Sherkatghanad:2014hda,Altamirano:2014tva,Kubiznak:2015bya,Hennigar:2015esa,Hendi:2015hoa,Hennigar:2015wxa,Hendi:2015cqz,Hennigar:2016xwd,Hendi:2016usw,Hendi:2017fxp,Jafarzade:2017kin,EslamPanah:2019szt} or the detailed review~\cite{Kubiznak:2016qmn}.  

Recently, it has been found that the horizon radius can be replaced by the shadow radius to study the BH thermodynamics. {Zhang et al. first used the shadow radius to reveal the phase structure of BHs~\cite{Zhang:2019glo}. They found that the phase structure of the spherically symmetric BH can be reflected by the shadow radius. For the axially symmetric BH, the phase structure can be reflected by the size of its shadow, but the distortion of the shadow does not reflect the phase structure.} Then the critical behavior of the Reissner–Nordström-AdS BH is investigated by the BH shadow in Ref.~\cite{Belhaj:2020nqy}. Fairly recently, the microstructure states and thermal profiles of different kinds of AdS BHs were studied using shadow formalism~\cite{Cai:2021uov,Wang:2021vbn,Guo:2022yjc,Du:2022hom}. Given the significance of BH thermodynamics and inspired by these pioneering works, we study the thermodynamics of the Hayward-AdS BH with the shadow.

The motivation for studying the Hayward-AdS BH is as follows. It is well known that Bardeen first proposed the regular BH model, which avoids the BH singularity~\cite{Bardeen1968}. Ayon-Beato and Garcia found that the physical source of the Bardeen BH is a nonlinear magnetic monopole~\cite{Ayon-Beato:1998hmi,Ayon-Beato:1999qin,Ayon-Beato:1999kuh,Ayon-Beato:2000mjt}. As another well-known regular BH, the Hayward BH was found in Ref.~\cite{Hayward:2005gi}. Then, it was generalized to the AdS spacetime and the critical phenomena of the Hayward-AdS BH was studied~\cite{Fan:2016rih}. For such a significant regular BH, we seek to establish a relation between the thermodynamic PT and the BH shadow. The impact of the magnetic charge $g$ on the thermodynamics of the Hayward-AdS BH is also worth our study.

The paper is organized as follows. A brief review on deriving the shadow radius of the Hayward-AdS BH is given in Sec.~\ref{sec:2}. In Sec.~\ref{sec:3}, the thermodynamic PT of the Hayward-AdS BH is investigated using shadow formalism. The effect of the magnetic charge on the thermodynamics is also discussed. In Sec.~\ref{sec:4}, we use the thermal profile of the BH shadow to clarify the thermodynamic PT of the Hayward-AdS BH. The relation between the thermodynamic PT and the BH shadow is established. The conclusions are given in Sec.~\ref{sec:5}. In this paper, we set the units $G_N=\hbar=\kappa_B=c=1$.

\section{Shadow of the Hayward-AdS BH}\label{sec:2}
In this section, we review the derivation of the shadow radius for the Hayward-AdS BH. The metric of a static spherically symmetric BH can be written as
\begin{equation}
	\label{2-1}
	d s^{2}=-f(r) d t^{2}+h(r)^{-1} d r^{2}+r^{2} (d \theta^{2}+  \sin ^{2} \theta d \phi^{2}),
\end{equation}
where $f(r)$ and $h(r)$ represent the lapse functions associated with the radius parameter $r$. For a photon moving in such spacetime background of the metric (\ref{2-1}), the corresponding Hamiltonian satisfies
\begin{equation}
\label{2-3}
2\mathcal{H}= g^{\mu \nu} p_{\mu} p_{\nu}=0,
\end{equation}
where $p_\mu=\frac{\mathrm{d} x_\mu}{\mathrm{d} \lambda}$ is the proper four-momentum of the photon and $\lambda$ denotes the affine parameter. For convenience, we assume that the photon is moving on the equatorial plane and therefore we can fix the coordinate $\theta=\pi/2$ and $\dot{\theta}=0$. Then, with Eq.~(\ref{2-3}) and the metric (\ref{2-1}), one can obtain 
\begin{equation}
	\label{2-4}
	-\frac{p_{t}^{2}}{2f(r)}+\frac{h(r) p_{r}^{2}}{2}+\frac{p_{\phi}^{2}}{2r^{2}}=0,
\end{equation}
where $p_r$ is the radial momentum of the photon. For the geodesic motion in the context of such BH background, there are two conserved quantities corresponding to the Killing vector fields $\frac{\partial}{\partial t}$ and $\frac{\partial}{\partial \phi}$, i.e.,
\begin{equation}
	-E =p_{t}=\frac{\partial \mathcal{H}}{\partial \dot{t}}, \quad
	L =p_{\phi} = \frac{\partial \mathcal{H}}{\partial \dot{\phi}},
\end{equation}
where $E=-p_{t}$ represents the energy of the photon and $L=p_{\phi}$ represents the angular momentum of the photon. Note that the dot indicates the derivative with respect to the affine parameter $\lambda$. Then, one can get the equation of motion of the photon from Eq.~(\ref{2-4}), 
\begin{eqnarray}\label{2-55}
	\label{2-5-1}
	\dot{t}=\frac{\partial \mathcal{H}}{\partial p_{t}}=-\frac{p_{t}}{f(r)}, \quad
	\label{2-5-2}
	\dot{\phi}=\frac{\partial \mathcal{H}}{\partial p_{\phi}}=\frac{p_{\phi}}{r^{2}}, \quad
	\label{2-5-3}
	\dot{r}=\frac{\partial \mathcal{H}}{\partial p_{r}}=p_{r} h(r).
\end{eqnarray}
From Eq.~(\ref{2-3}), the effective potential of the photon can be defined as
\begin{equation}
	\label{2-6}
	\dot{r}^{2} + V_{eff}(r) = 0.
\end{equation}
With Eqs.~(\ref{2-4}) and (\ref{2-55}), the effective potential $V_{ eff}$ can be written as 
\begin{equation}
	\label{2-7}
	V_{ eff}(r) = h(r)\Bigg(\frac{L^2}{r^2}-\frac{E^{2}}{f(r)}\Bigg).
\end{equation}
The circular orbit radius $r_p$ of the photon needs to satisfy the following conditions,  
\begin{equation}
	\label{2-8}
	V_{eff}\left(r_p\right)=0,\left.\quad \frac{\partial V_{eff}(r)}{\partial r}\right|_{r=r_p}=0.
\end{equation}
The impact parameter $\mu_{p}$ of the photon can be deduced from Eqs.~(\ref{2-7}) and (\ref{2-8}), which is given by
\begin{equation}\label{2-88}
	\mu_{p} = \frac{L}{E}=\left.\frac{r}{\sqrt{f(r)}}\right|_{r=r_{p}}.
\end{equation}
Using Eq.~(\ref{2-55}), the orbit equation of the photon can be expressed as 
\begin{equation}
	\label{2-99}
	\frac{d r}{d \phi}=\frac{\dot{r}}{\dot{\phi}}=\frac{r^{2} h(r) p_{r}}{L}.
\end{equation}
Combining with Eqs.~(\ref{2-4}) and (\ref{2-99}), one can obtain
\begin{equation}
	\label{2-9}
	\frac{{ d} r}{{ d} \phi} = \pm r \sqrt{h(r)\left(\frac{r^{2}E^{2}}{f(r)L^{2}}-1\right)}.
\end{equation}
For the turning point ($r=\chi$) of the photon orbit, one has the mathematical constraint $\frac{{ d} r}{{ d} \phi}{\big |}_{r=\chi}=0$, which means that $\frac{E^2}{L^2}=\frac{f(\chi)}{\chi^2}$. Then Eq.~(\ref{2-9}) can be rewritten as
\begin{equation}
	\label{2-10}
	\frac{{ d} r}{{ d} \phi} = \pm r \sqrt{h(r)\left(\frac{r^{2}f(\chi)}{f(r)\chi^{2}}-1\right)}.
\end{equation}

Now, we study the BH shadow observed by a static observer located at position $r_o$. For a light ray emitting from the observer and transmitting into the past with an angular $\psi$ with respect to the radial direction, we have~\cite{Zhang:2019glo,Belhaj:2020nqy,Cai:2021uov,Wang:2021vbn},
\begin{equation}
	\label{2-11}
	\cot \psi = \frac{\sqrt{g_{r r}}}{\sqrt{g_{\phi \phi}}} \cdot \frac{d r}{d \phi}{\Bigg |}_{r=r_{o}} = \left.\frac{1}{r \sqrt{h(r)}} \cdot \frac{d r}{d \phi}\right|_{r=r_o}.
\end{equation}
Combining with Eqs.~(\ref{2-10}) and (\ref{2-11}), we get
\begin{equation}
	\label{2-122}
	\cot ^2 \psi=\frac{r_o^2 f(\chi)}{f\left(r_o\right) \chi^2}-1 \quad {\text{or}}  \quad \sin ^2 \psi=\frac{f\left(r_o\right) \chi^2}{r_o^2 f(\chi)}.
\end{equation}
For the static observer located at position $r_o$, one can set $\chi = r_p$ to obtain the shadow radius $r_{s}$,
\begin{equation}
	\label{2-13}
	r_{s}=  r_{o} \sin \psi = \chi\sqrt{\frac{f(r_{o})}{f(\chi)}}{\Bigg |}_{\chi = r_{p}}.
\end{equation}

For the Hayward-AdS BH, the coefficients in the metric (\ref{2-1}) are given as~\cite{Hayward:2005gi,Fan:2016rih}
\begin{equation}
\label{2-14}
f(r)=h(r)^{-1}=1-\frac{2 M r^{2}}{r^{3}+g^{3}}+\frac{8 \pi P r^{2}}{3},
\end{equation}
where $M$ represents the BH mass and $g$ represents the magnetic charge. The thermodynamic pressure $P$ is associated with the cosmological constant, namely, $P =-\frac{\Lambda}{8\pi} =\frac{3}{8 \pi l^{2}}$. 
One can use Eqs.~(\ref{2-8}) and (\ref{2-14}) to obtain the circular orbit radius of the photon, which is given as
\begin{equation}
\label{2-18}
r_{p}=M+\frac{M^{2}}{\left(M^{3}-g^{3}+\sqrt{g^{6}-2 g^{3} M^{3}}\right)^{1 / 3}}+\left(M^{3}-g^{3}+\sqrt{g^{6}-2 g^{3} M^{3}}\right)^{1 / 3}.
\end{equation}
Finally, with Eqs.~(\ref{2-13}) and (\ref{2-14}), the expression for the shadow radius of the Hayward-AdS BH, for a static observer located at position $r_o$, can be directly written as
\begin{equation}
\label{2-19}
r_{s} = r_{p} \sqrt{\frac{f(r_{o})}{f(r_{ p})}},
\end{equation}
where $r_p$ is given by Eq.~(\ref{2-18}).
\section{Thermodynamics of the Hayward-AdS BH using shadow formalism}\label{sec:3}

For the Hayward-AdS BH, the mass $M$ and temperature $T$ are given by~\cite{Fan:2016rih}
\begin{equation}
	\label{2-15}
	M=\frac{g^{3}}{2 r_{h}^{2}}+\frac{r_{h}}{2}+\frac{4}{3} P \pi\left(r_{h}^{3}+g^{3}\right), \quad T=\frac{r_{h}^{3}-2 g^{3}}{4 \pi r_{h}\left(g^{3}+r_{h}^{3}\right)}+\frac{2 P r_{h}^{4}}{g^{3}+r_{h}^{3}},
\end{equation}
where $r_{h}$ is the horizon radius of the Hayward-AdS BH. Since the expression of $r_{h}$ is extremely verbose, we do not display the exact form here. The equation of state of the Hayward-AdS BH is given as 
\begin{equation}
	\label{2-17}
	P=\frac{g^{3}}{4 \pi r_{h}^{5}}-\frac{1}{8 \pi r_{h}^{2}}+\frac{g^{3} T}{2 r_{h}^{4}}+\frac{T}{2 r_{h}}.
\end{equation}
To obtain the critical point of the thermodynamic PT for the Hayward-AdS BH, one can take Eq.~(\ref{2-17}) into the critical condition $({\partial P}/{\partial r_{ h}})=0=({{\partial}^{2}P}/{\partial r_{ h}^{2}})$, which yields
\begin{eqnarray}
\label{3-1-1}
&&P_{ c} =\frac{3(57-23 \sqrt{6})(14+6 \sqrt{6})^{1 / 3}}{800 g^{2} \pi} \simeq 0.002418 g^{-2},\\
\label{3-1-2}
&&r_{ c}=(14+6 \sqrt{6})^{1 / 3} g \simeq 3.061577 g,\\
\label{3-1-3}
&&T_{ c}=\frac{(5-2 \sqrt{6})(7+3 \sqrt{6})^{2 / 3}}{4 \times 2^{1 / 3} g \pi}\simeq 0.037675 g^{-1}.
\end{eqnarray}
Substituting Eq.~(\ref{2-15}) into Eq.~(\ref{2-18}), one can rewrite the circular orbit radius of the photon as
\begin{equation}
	\label{rsrh}
	r_{ p}=\frac{g^3}{2 r_{ h}^2}+\frac{r_{ h}}{2}+\frac{4}{3} \left(g^3+r_{ h}^3\right)P \pi+\frac{\left(3+8 P \pi r_{ h}^2\right)^2\left(g^3+r_{ h}^3\right)^2}{36 r_{ h}^4 \xi^{1 / 3}}+\xi^{1 / 3},
\end{equation}
where
\begin{equation}
	\label{xi}
	\xi=\frac{\left(3+8 P \pi r_{ h}^2\right)^3\left(g^3+r_{ h}^3\right)^3}{216 r_{ h}^6}+\sqrt{g^6-\frac{g^3\left(3+8 P \pi r_{ h}^2\right)^3\left(g^3+r_{ h}^3\right)^3}{108 r_{ h}^6}}-g^3.
\end{equation}

Combining with Eqs.~(\ref{2-19}) and (\ref{rsrh}), we plot the $r_{ s}-r_{ h}$ diagrams of the Hayward-AdS BH in Fig.~\ref{fig1}, where we have set $f(r_{o}=100)=1$ for the static observer\cite{Zhang:2019glo}. For different magnetic charge $g$ and pressure $P$, the shadow radius $r_{ s}$ invariably increases monotonically with the horizon radius $r_{h}$. This result implies that the horizon radius $r_{h}$ can be substituted with the shadow radius $r_{ s}$ to study the thermodynamics of the Hayward-AdS BH.

\begin{figure}[h]
	\centering

	\subfigure[$g=0.3$\label{ptr1}]{
		\begin{minipage}[t]{0.33\linewidth}
			\centering
			\includegraphics[width=2.5in]{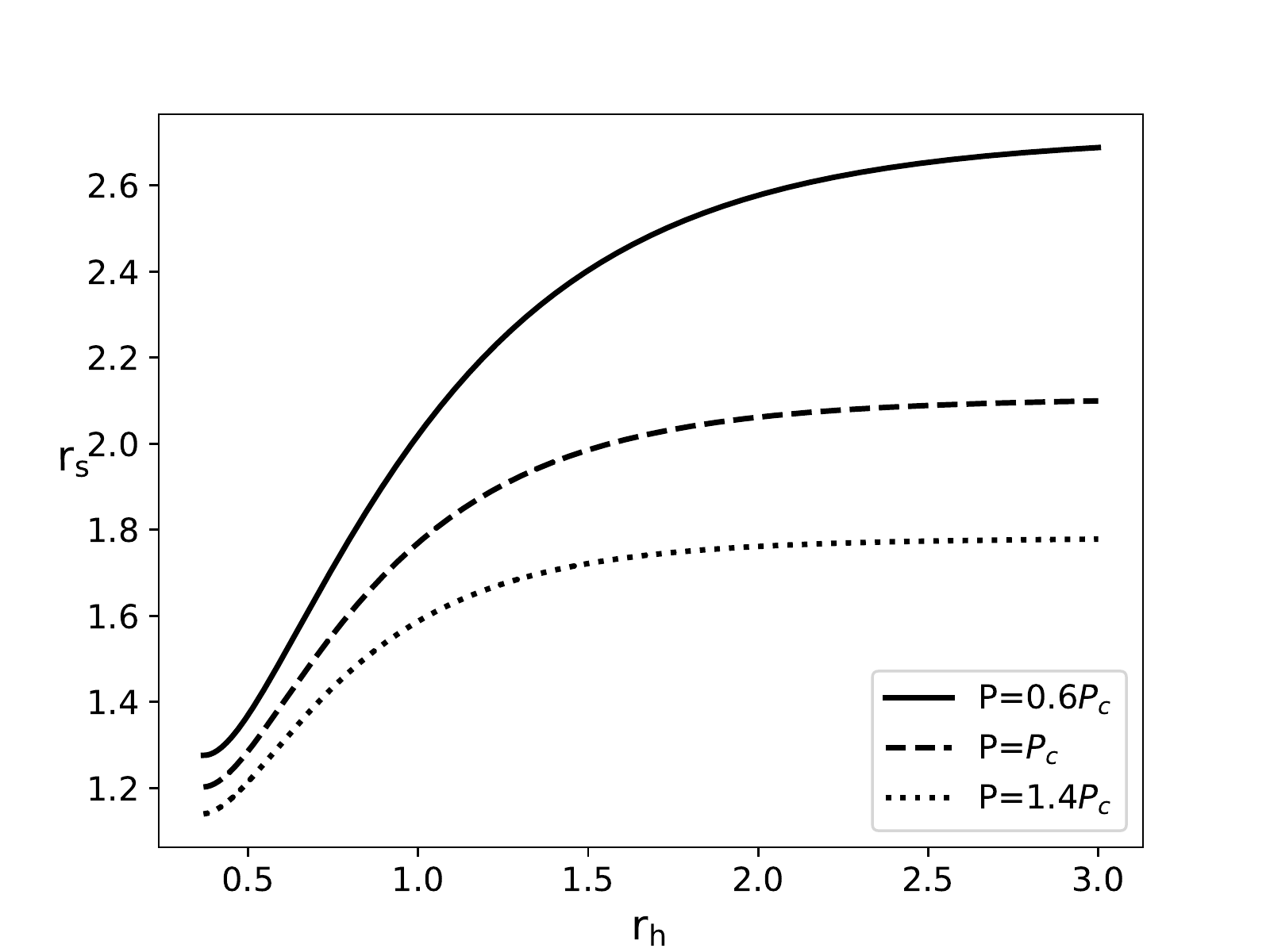}
			%\caption{fig1}
		\end{minipage}%
	}%
	\subfigure[$g=0.6$\label{ptr2}]{
		\begin{minipage}[t]{0.33\linewidth}
			\centering
			\includegraphics[width=2.5in]{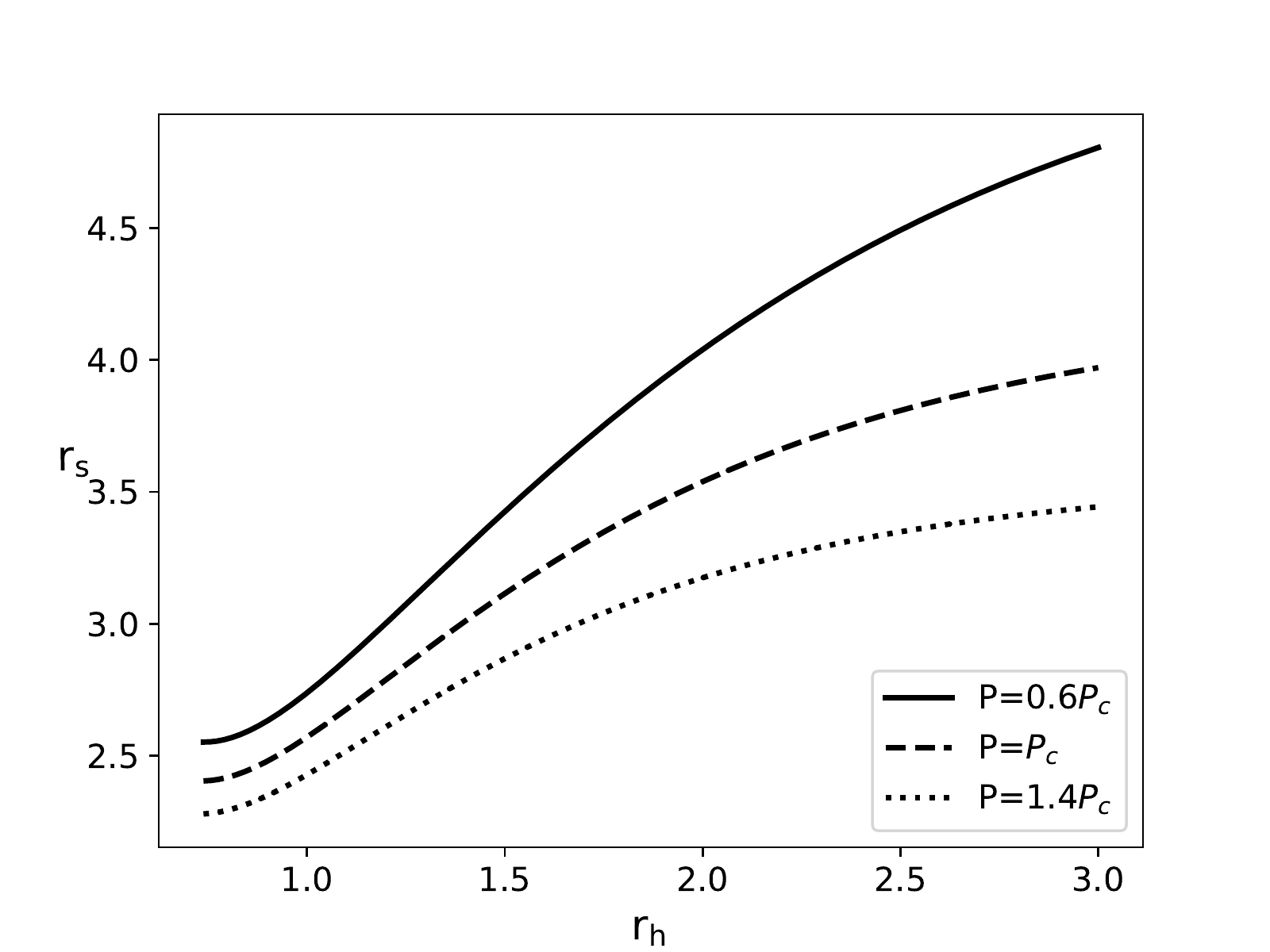}
			%\caption{fig2}
		\end{minipage}%
	}%
	\subfigure[$g=0.9$\label{ptr3}]{
		\begin{minipage}[t]{0.33\linewidth}
			\centering
			\includegraphics[width=2.5in]{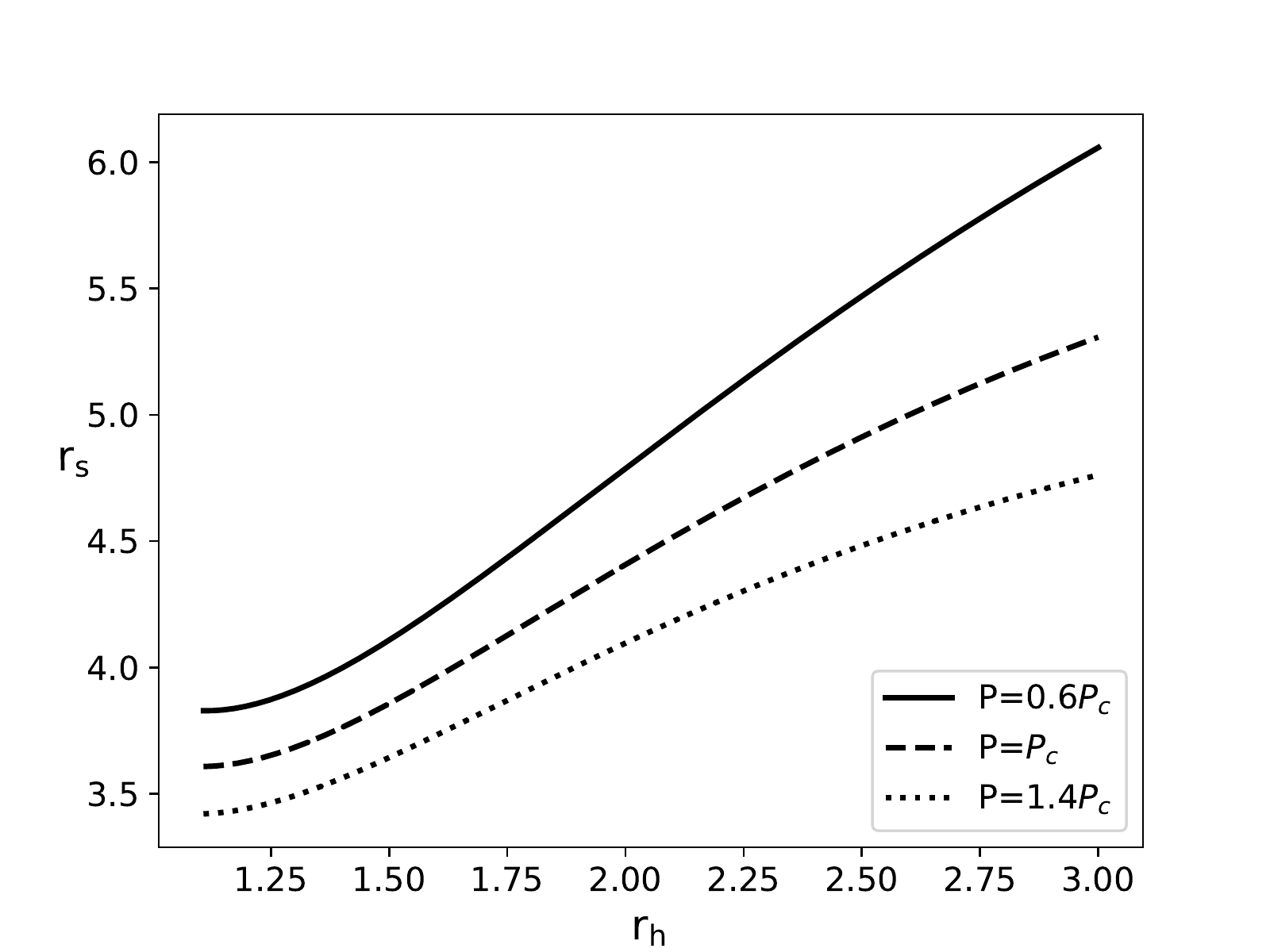}
			%\caption{fig2}
		\end{minipage}
	}%	
	
	\caption{The shadow radius $r_{ s}$ versus the horizon radius $r_{ h}$. The solid, dashed and dotted lines correspond to $P=0.6P_{c}$, $P=P_{c}$ and $P=1.4P_{c}$, respectively.}
	\label{fig1}
\end{figure}

With Eq.~(\ref{2-15}), we plot the $T-r_{ h}$ diagrams with three types of pressures based on the critical pressure $P_c$ (see Figs.~\ref{fig21}, \ref{fig22} and \ref{fig23} with the magnetic charge $g=0.3, 0.6, 0.9$). It is found that when the pressure is less than the critical pressure such as $P=0.6P_{ c}$, the corresponding temperature curve is not monotonous and there exists a vdW-like PT. The slopes of the small radius region ($r_{h}<r_{h1}$) and the large radius region ($r_{h}>r_{h2}$) are both positive, which correspond to the small BH phase and the large BH phase, respectively. The intermediate region ($r_{h1}<r_{h}<r_{h2}$) represents a coexistence region of the small and large BH. The intermediate region is thermodynamically unstable and the corresponding temperature $T_{co}$ is known as coexistence temperature. For the pressure equal to the critical pressure ($P=P_{ c}$), there will be an unstable PT. As for the pressure larger than the critical pressure ($P=1.4P_{ c}$), PT will not appear.

On the other hand, we can also study the thermodynamic PT of the Hayward-AdS BH by replacing the horizon radius $r_{h}$ with the shadow radius $r_{s}$. In Figs.~\ref{fig24}, \ref{fig25} and \ref{fig26}, we plot the $T-r_{s}$ diagrams with the parameters adopted in the $T-r_{h}$ diagrams. To observe the difference between the two kinds of diagrams, we also consider three sets of values of the magnetic charge $g$ and three sets of values of the pressure $P$. The results show that the PT process denoted by the shadow radius $r_s$ is similar to the PT process denoted by the horizon radius $r_h$. The critical point of the PT process is still dependent on the pressure $P$. For $P<P_c$, there exists a vdW-like PT. The small ($r_{ s}<r_{ s1}$) and large ($r_{ s}>r_{ s2}$) radius regions correspond to the stable small and large BH, respectively. The radius region $r_{ s1}<r_{s}<r_{ s2}$ corresponds to the unstable intermediate BH. For $P=P_c$, there will be an unstable PT. When $P>P_c$, there is no PT. Moreover, from the whole Fig.~\ref{fig2}, it can be seen that for the pressure $P=0.6P_c$, as the magnetic charge $g$ increases, the shadow radius $r_{s}$ becomes larger while the coexistence temperature $T_{co}$ becomes lower.
\begin{figure}[h]
	\centering

	\subfigure[$T-r_{ h}$ with $g=0.3$\label{fig21}]{
		\begin{minipage}[t]{0.32\linewidth}
			\centering
			\includegraphics[width=2.00in]{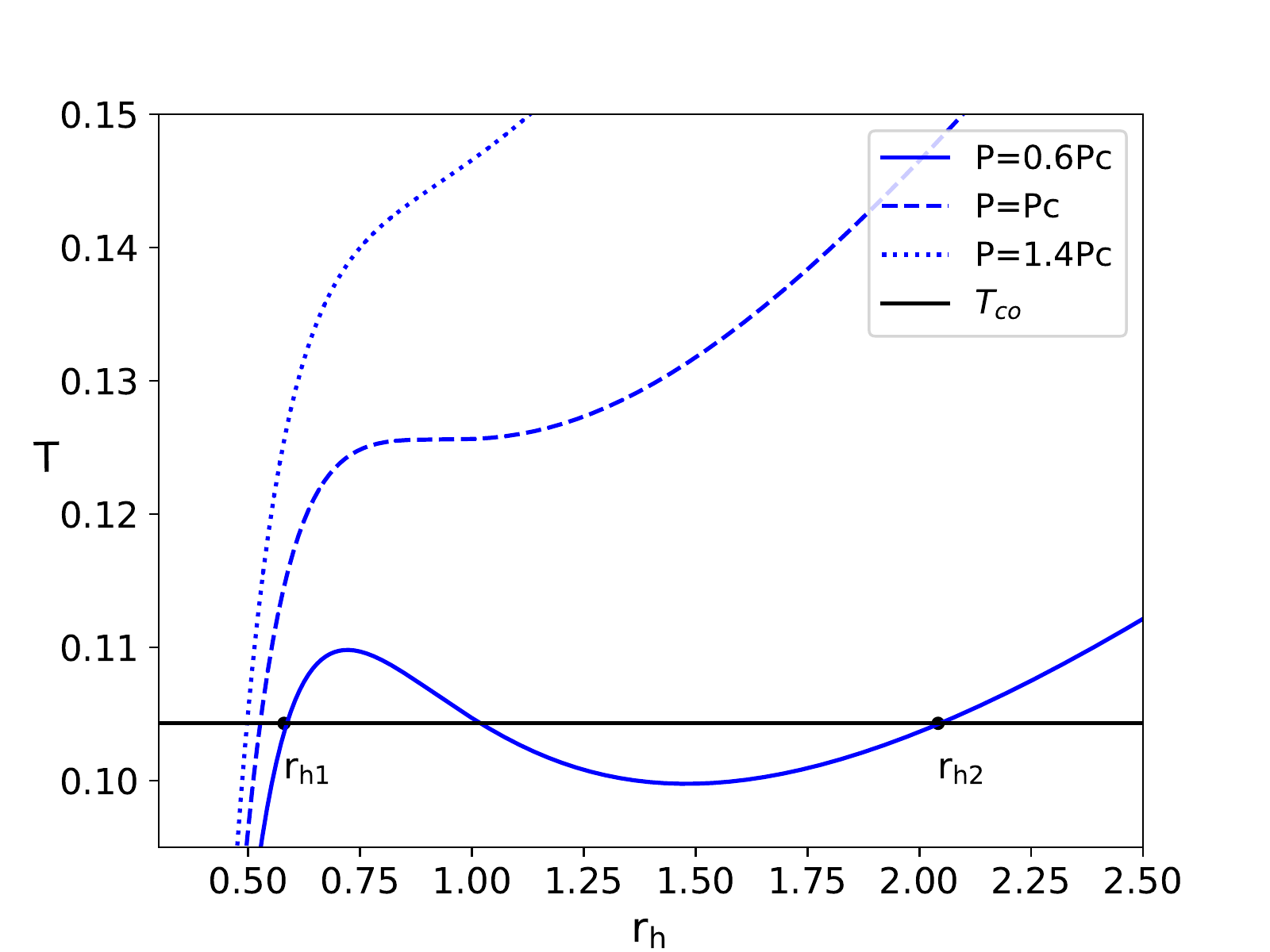}
			%\caption{fig1}
		\end{minipage}%
	}%
	\subfigure[$T-r_{ h}$ with $g=0.6$\label{fig22}]{
		\begin{minipage}[t]{0.32\linewidth}
			\centering
			\includegraphics[width=2.00in]{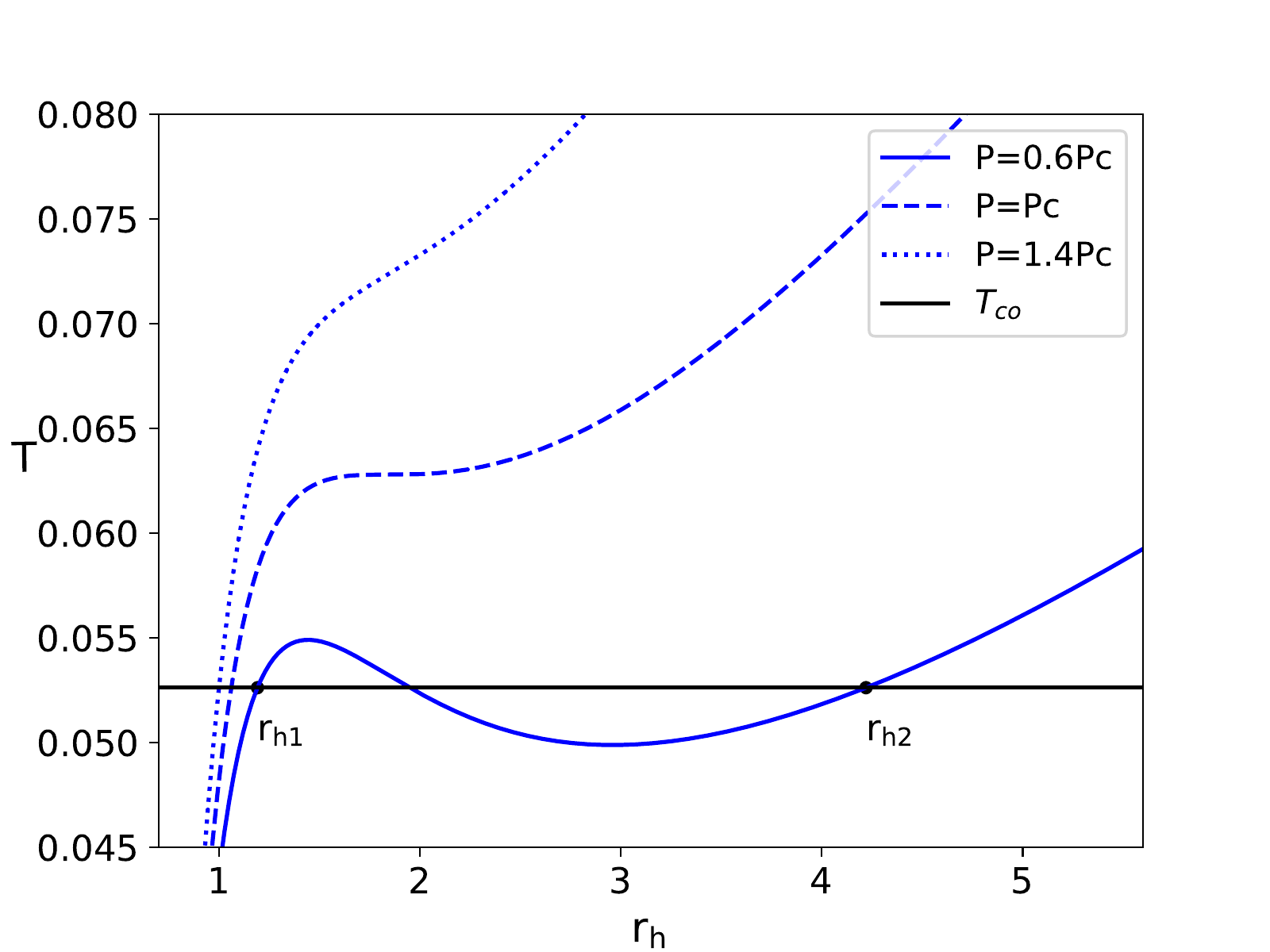}
			%\caption{fig2}
		\end{minipage}%
	}%
	\subfigure[$T-r_{ h}$ with $g=0.9$\label{fig23}]{
		\begin{minipage}[t]{0.32\linewidth}
			\centering
			\includegraphics[width=2.00in]{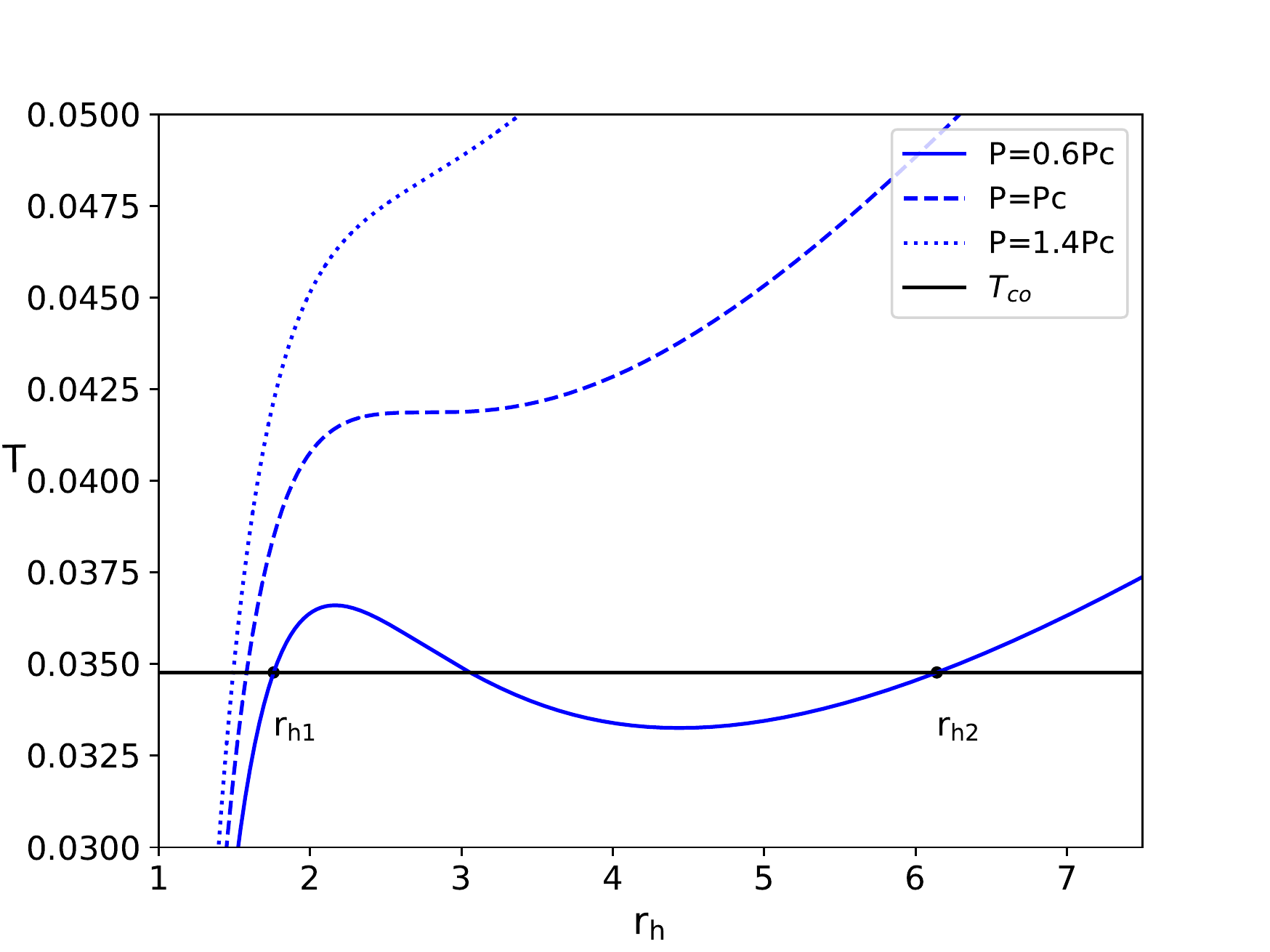}
			%\caption{fig2}
		\end{minipage}
	}%
	\quad
	\subfigure[$T-r_{ s}$ with $g=0.3$\label{fig24}]{
		\begin{minipage}[t]{0.32\linewidth}
			\centering
			\includegraphics[width=2.00in]{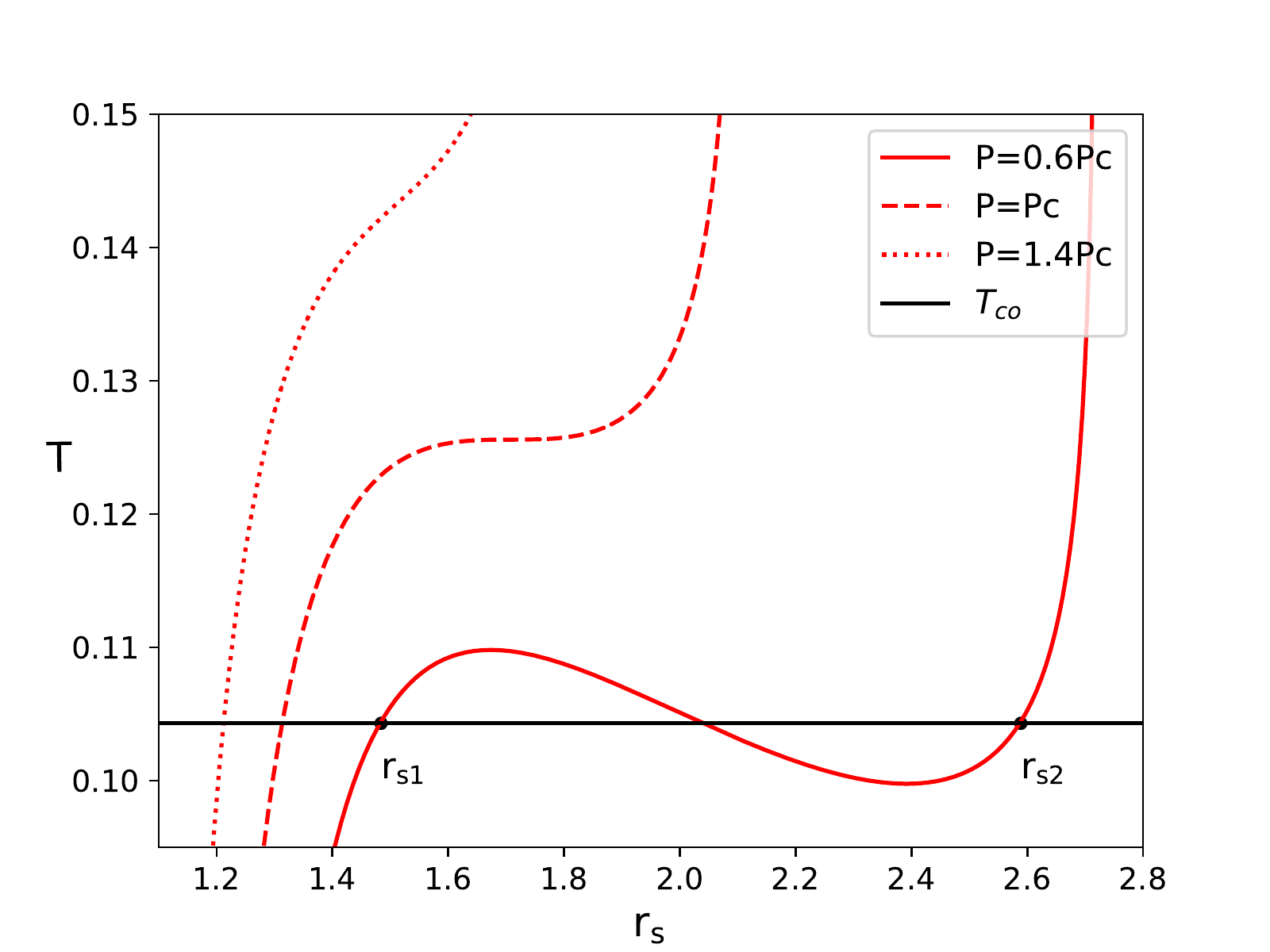}
			%\caption{fig2}
		\end{minipage}
	}%
	\subfigure[$T-r_{ s}$ with $g=0.6$\label{fig25}]{
		\begin{minipage}[t]{0.32\linewidth}
			\centering
			\includegraphics[width=2.00in]{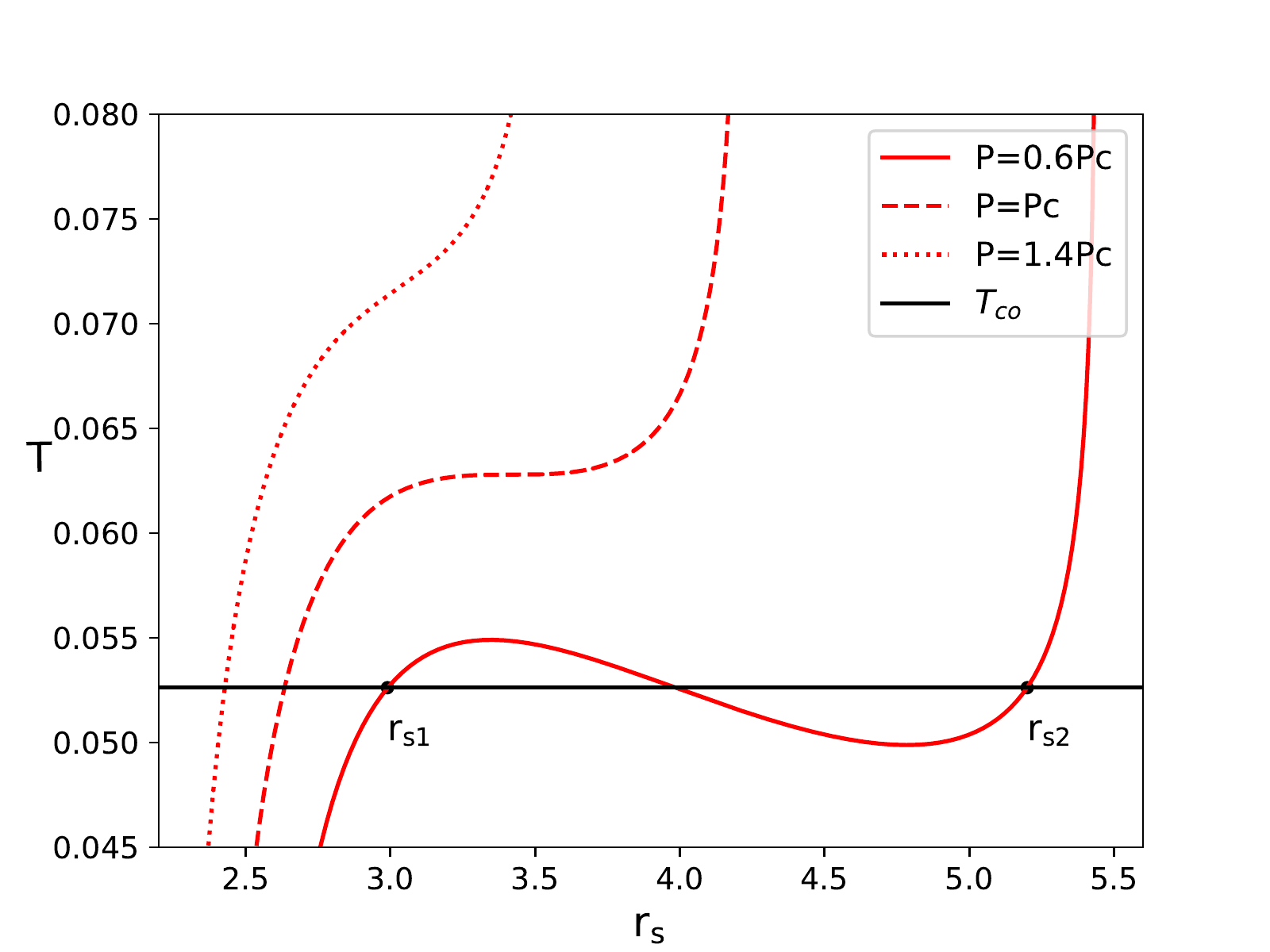}
			%\caption{fig2}
		\end{minipage}
	}%
	\subfigure[$T-r_{ s}$ with $g=0.9$\label{fig26}]{
		\begin{minipage}[t]{0.32\linewidth}
			\centering
			\includegraphics[width=2.00in]{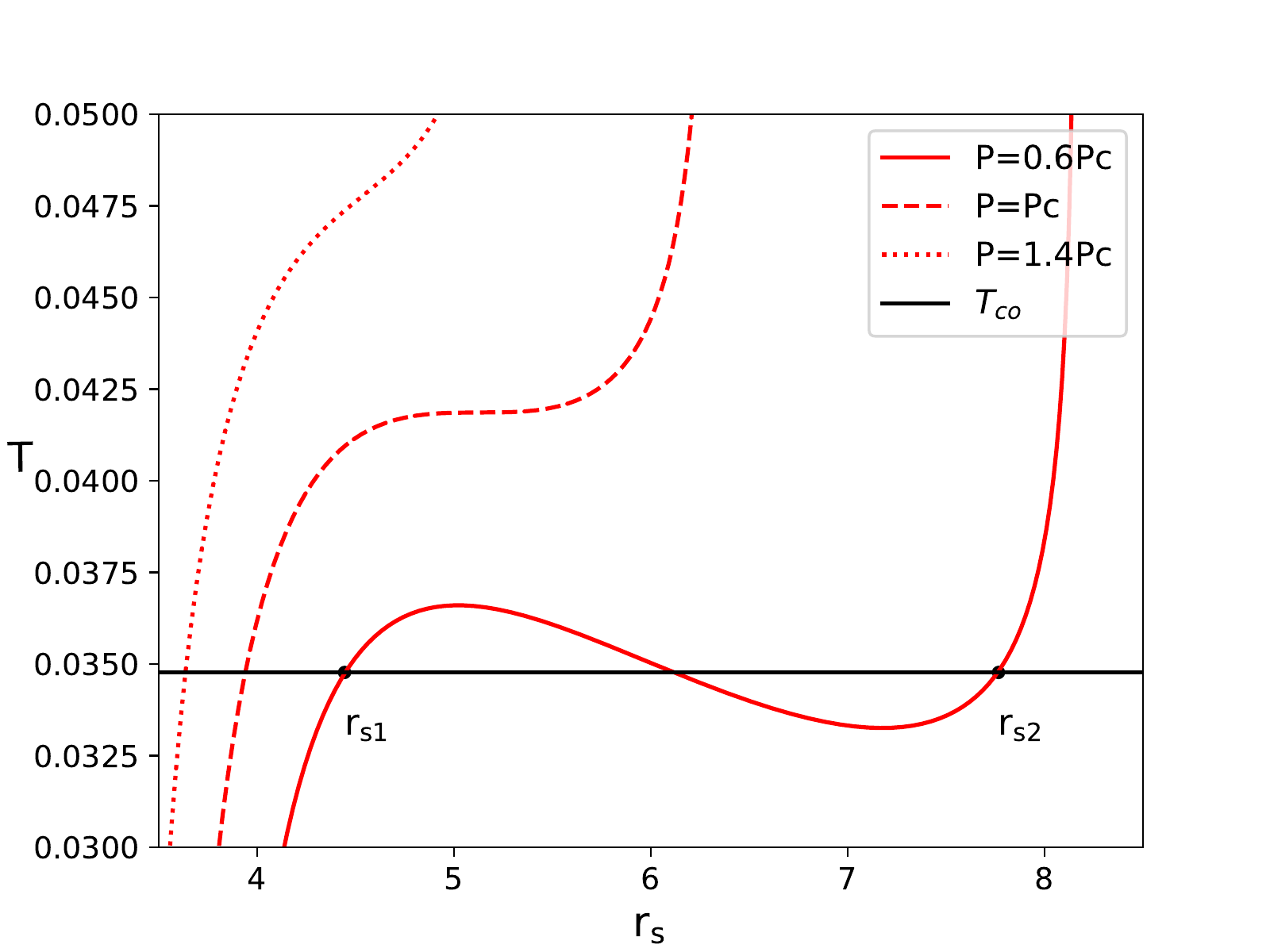}
			%\caption{fig2}
		\end{minipage}
	}%	
	
	\caption{The temperature $T$ versus $r_h$ (\ref{fig21}, ~\ref{fig22}, ~\ref{fig23}) and $r_{s}$ (\ref{fig24},~\ref{fig25},~\ref{fig26}) for the magnetic charge $g = 0.3,~0.6,~0.9$. $T_{co}$ (black lines) is the coexistence temperature. The solid, dashed and dotted lines correspond to $P=0.6P_{c}$, $P=P_{c}$ and $P=1.4P_{c}$, respectively.}
	\label{fig2}
\end{figure}

In order to analyze further the thermodynamic stability of the Hayward-AdS BH, we investigate its heat capacity. The positive (negative) heat capacity corresponds to the thermodynamically stable (unstable) BH phase. The heat capacity $C_P$ of the Hayward-AdS BH with a fixed pressure $P$ and magnetic charge $g$ is given by
\begin{equation}
\label{3-3}
C_{ P}=T\Bigg(\frac{dS}{dT}\Bigg)_{ P,g}=\frac{2 \pi r_{h}^{2}\left(g^{3}+r_{h}^{3}\right)\left(-2 g^{3}+r_{h}^{3}+8 P \pi r_{h}^{5}\right)}{2 g^{6}+r_{h}^{6}\left(-1+8 P \pi r_{h}^{2}\right)+2 g^{3} r_{h}^{3}\left(5+16 P \pi r_{h}^{2}\right)}.
\end{equation}

The $C_{P}-r_{h}$ diagrams of the Hayward-AdS BH for different magnetic charge $g$ and pressure $P$ are shown in Figs.~\ref{fig31}, \ref{fig32} and \ref{fig33}. For the pressure $P=0.6P_c$, $\left(\partial_{r_{h}} T\right)_P=0$ has two roots $r_{h1}$ and $r_{h2}$ as shown in Fig.~\ref{fig2}. These two roots indicate two divergences of the heat capacity $C_P$. The sign of $C_P$ is changed at these two divergences. When the pressure is increased to $P=P_c$, these two roots coincide with each other, so there is only one point of divergence in the heat capacity $C_P$. For the pressure $P=1.4P_c$, it can be seen that the heat capacity $C_P$ is a positive and continuous function of the horizon radius $r_{h}$. 
	
Combining with Eqs.~(\ref{2-19}) and (\ref{3-3}), we can also study the heat capacity of the Hayward-AdS BH by replacing the horizon radius $r_{h}$ with the shadow radius $r_{s}$. The $C_{P}-r_{s}$ diagrams of the Hayward-AdS BH are shown in Figs.~\ref{fig34}, \ref{fig35} and \ref{fig36}. When the pressure is less than the critical pressure, the heat capacity $C_P$ also diverges at two points. When the pressure is equal to the critical pressure, there is only one divergent behavior in the heat capacity $C_P$. For the pressure larger than the critical pressure, the heat capacity $C_P$ is also a positive and continuous function of the shadow radius $r_{s}$. 

To sum up,  the behaviors of the heat capacity $C_P$ in the $C_{P}-r_{s}$ diagrams are similar to those in the $C_{P}-r_{h}$ diagrams. In addition, Fig.~\ref{fig2} shows that for $P<P_c$, both the small ($r_{ s}<r_{ s1}$) and large ($r_{ s}>r_{ s2}$) BH branches have positive slopes (i.e., positive heat capacities), which means that these two branches are thermodynamically stable. The slope of the intermediate BH branch is negative, which represents a negative heat capacity, and thus it is thermodynamically unstable. These results are consistent with the phenomena depicted in Fig.~\ref{fig3}.

\begin{figure}[h]
	\centering

	\subfigure[$C_P-r_{ h}$ with $g=0.3$\label{fig31}]{
		\begin{minipage}[t]{0.32\linewidth}
			\centering
			\includegraphics[width=2.00in]{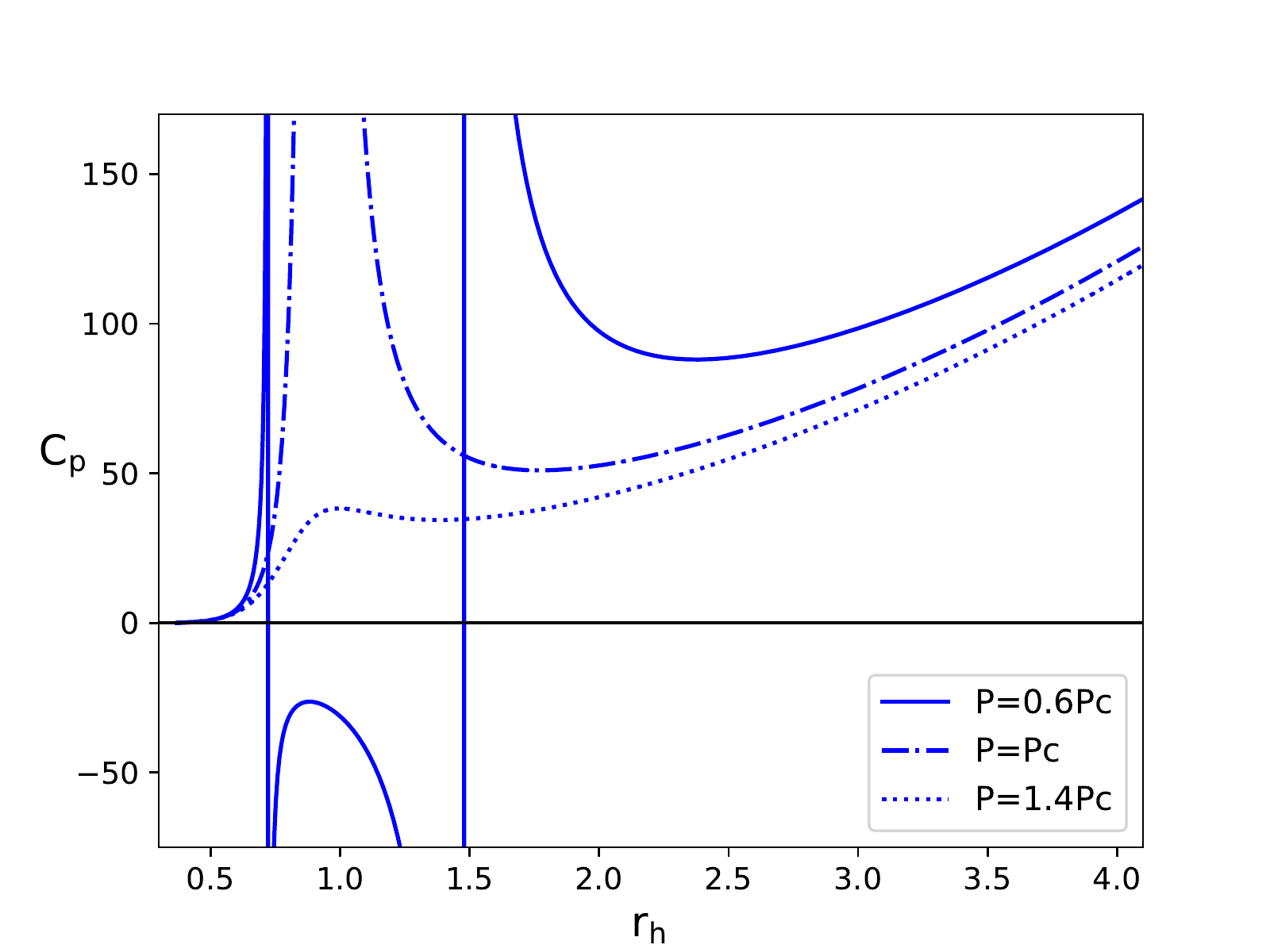}
			%\caption{fig1}
		\end{minipage}%
	}%
	\subfigure[$C_P-r_{ h}$ with $g=0.6$\label{fig32}]{
		\begin{minipage}[t]{0.32\linewidth}
			\centering
			\includegraphics[width=2.00in]{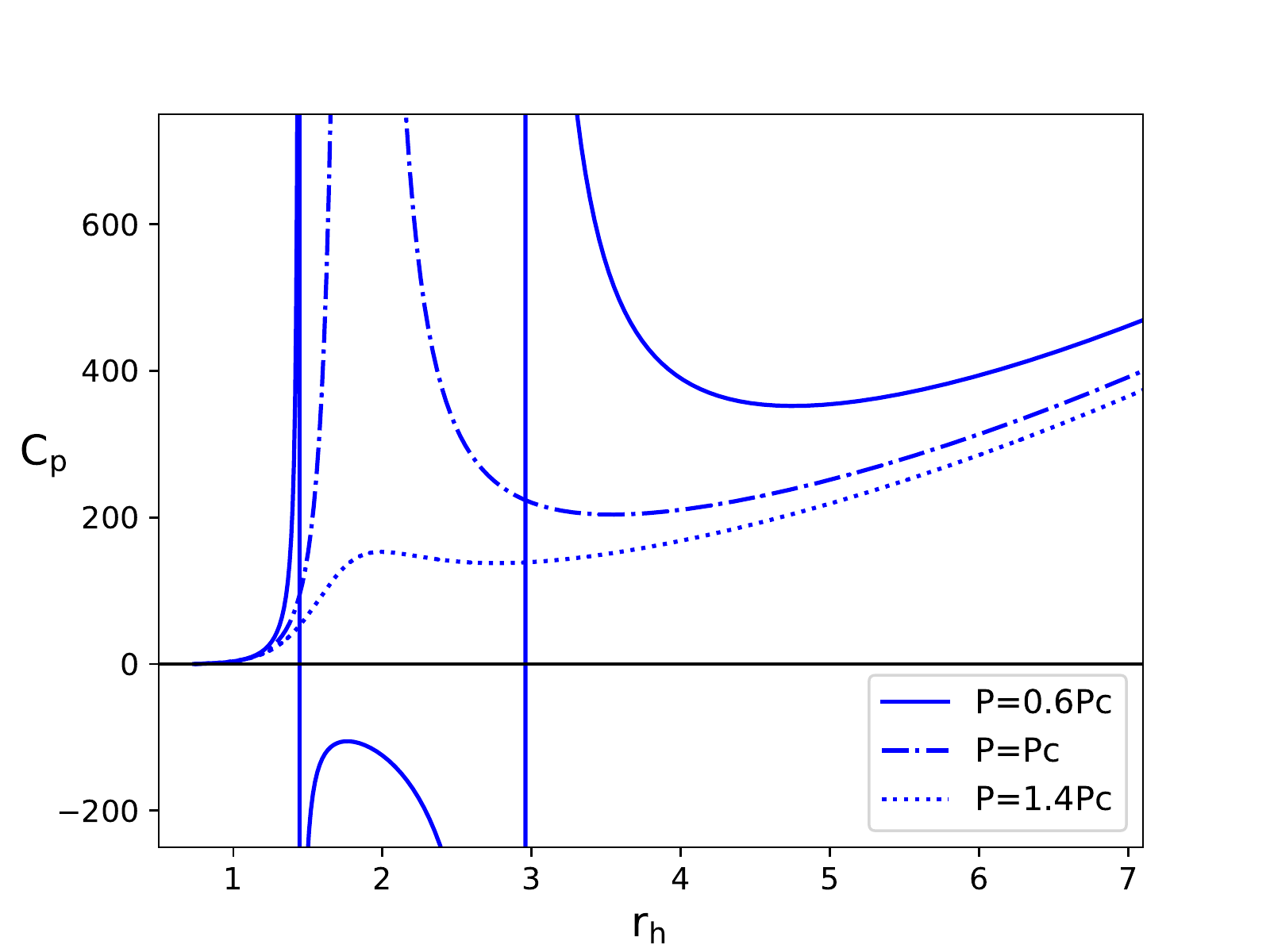}
			%\caption{fig2}
		\end{minipage}%
	}%
	\subfigure[$C_P-r_{ h}$ with $g=0.9$\label{fig33}]{
		\begin{minipage}[t]{0.32\linewidth}
			\centering
			\includegraphics[width=2.00in]{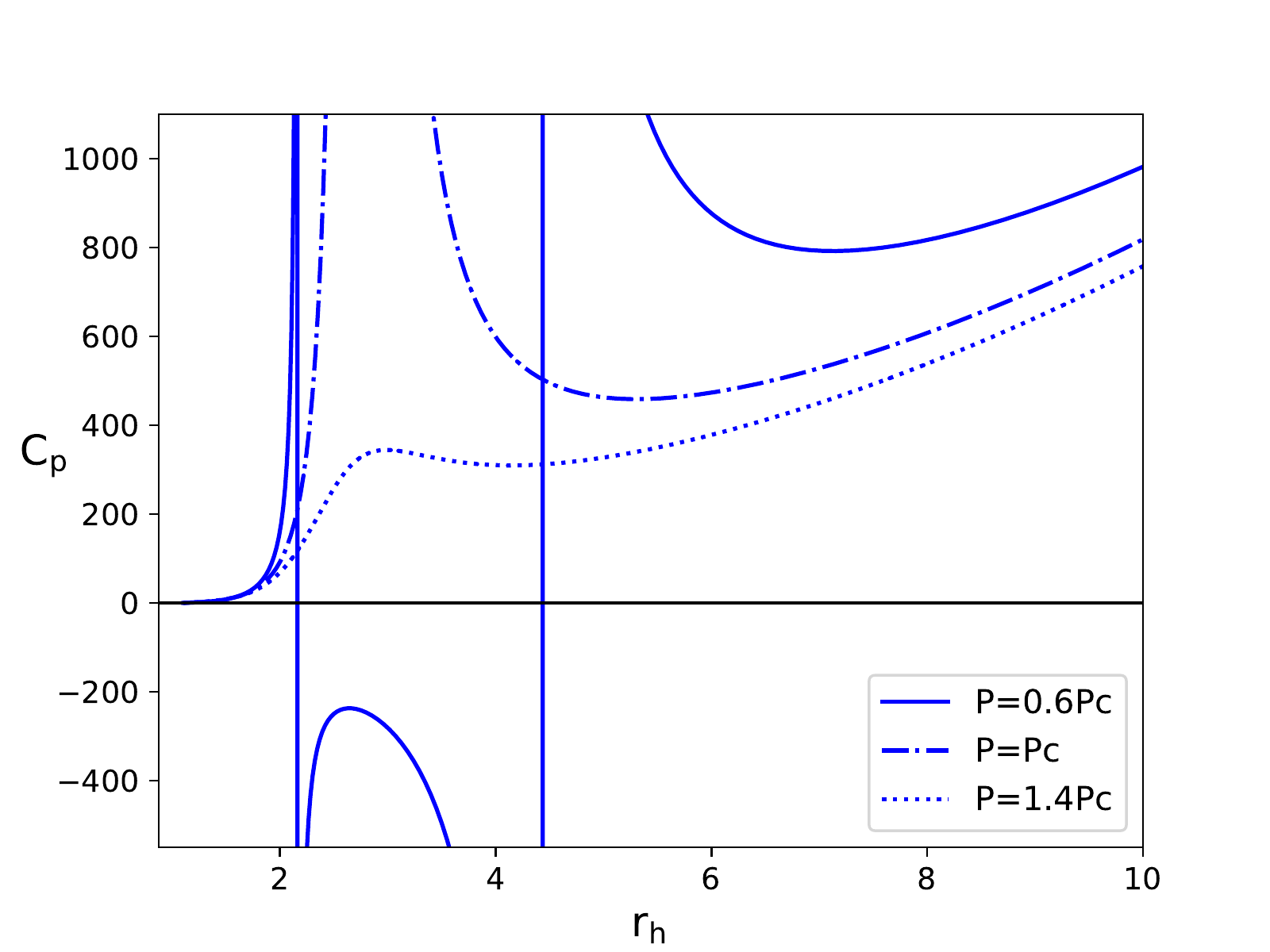}
			%\caption{fig2}
		\end{minipage}
	}%
	\quad
	\subfigure[$C_P-r_{ s}$ with $g=0.3$\label{fig34}]{
		\begin{minipage}[t]{0.32\linewidth}
			\centering
			\includegraphics[width=2.00in]{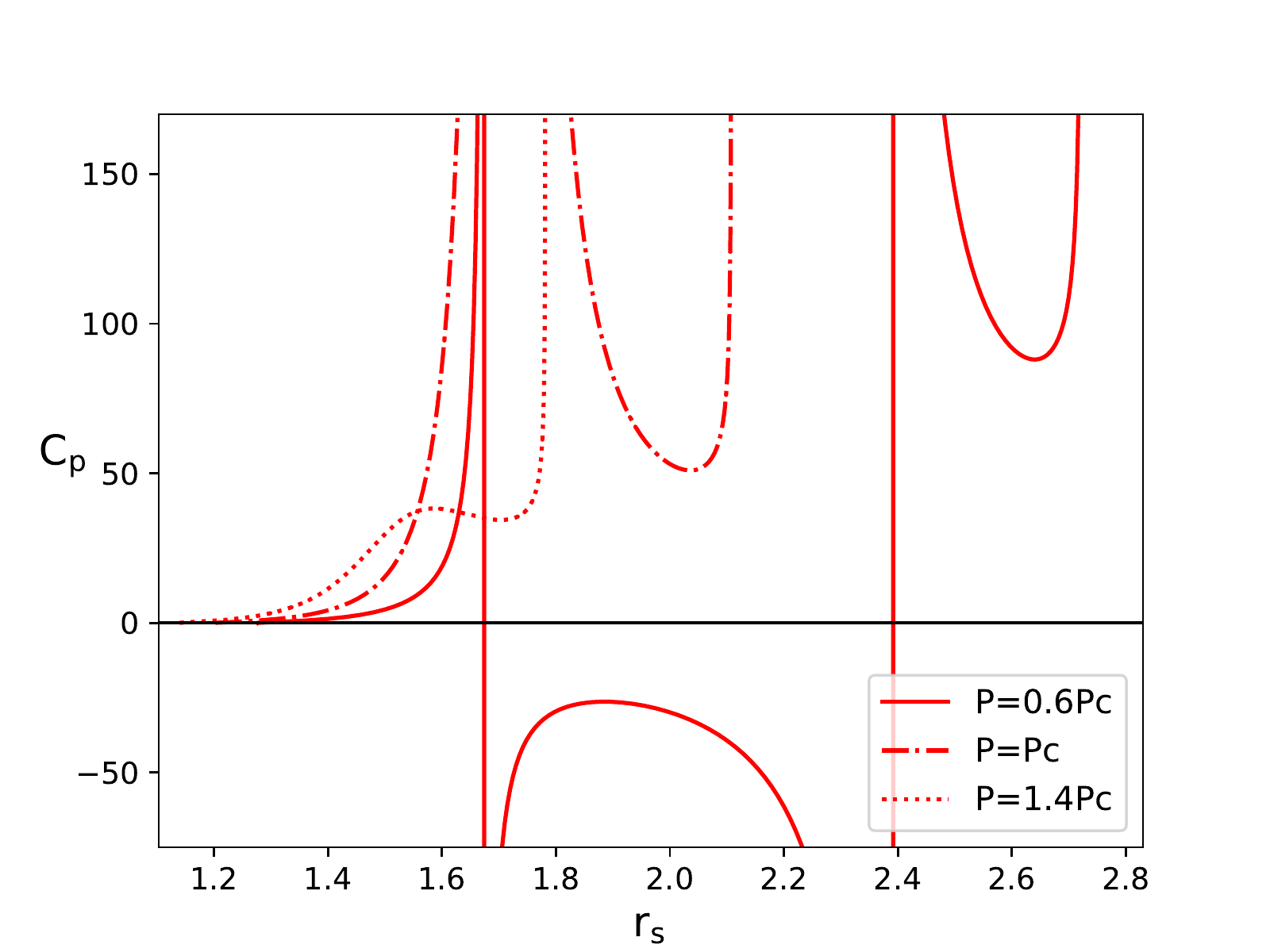}
			%\caption{fig2}
		\end{minipage}
	}%
	\subfigure[$C_P-r_{ s}$ with $g=0.6$\label{fig35}]{
		\begin{minipage}[t]{0.32\linewidth}
			\centering
			\includegraphics[width=2.00in]{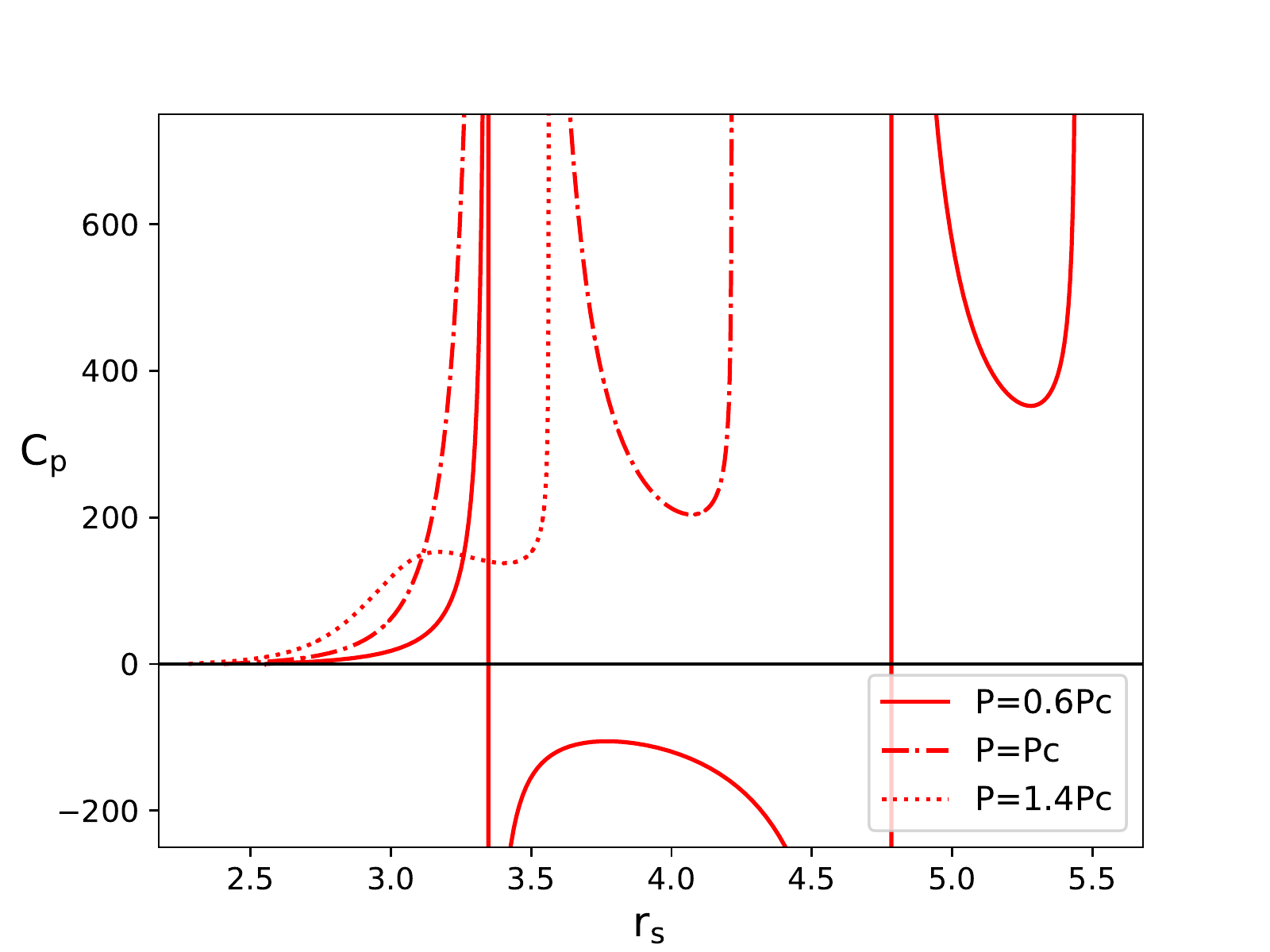}
			%\caption{fig2}
		\end{minipage}
	}%
	\subfigure[$C_P-r_{ s}$ with $g=0.9$\label{fig36}]{
		\begin{minipage}[t]{0.32\linewidth}
			\centering
			\includegraphics[width=2.00in]{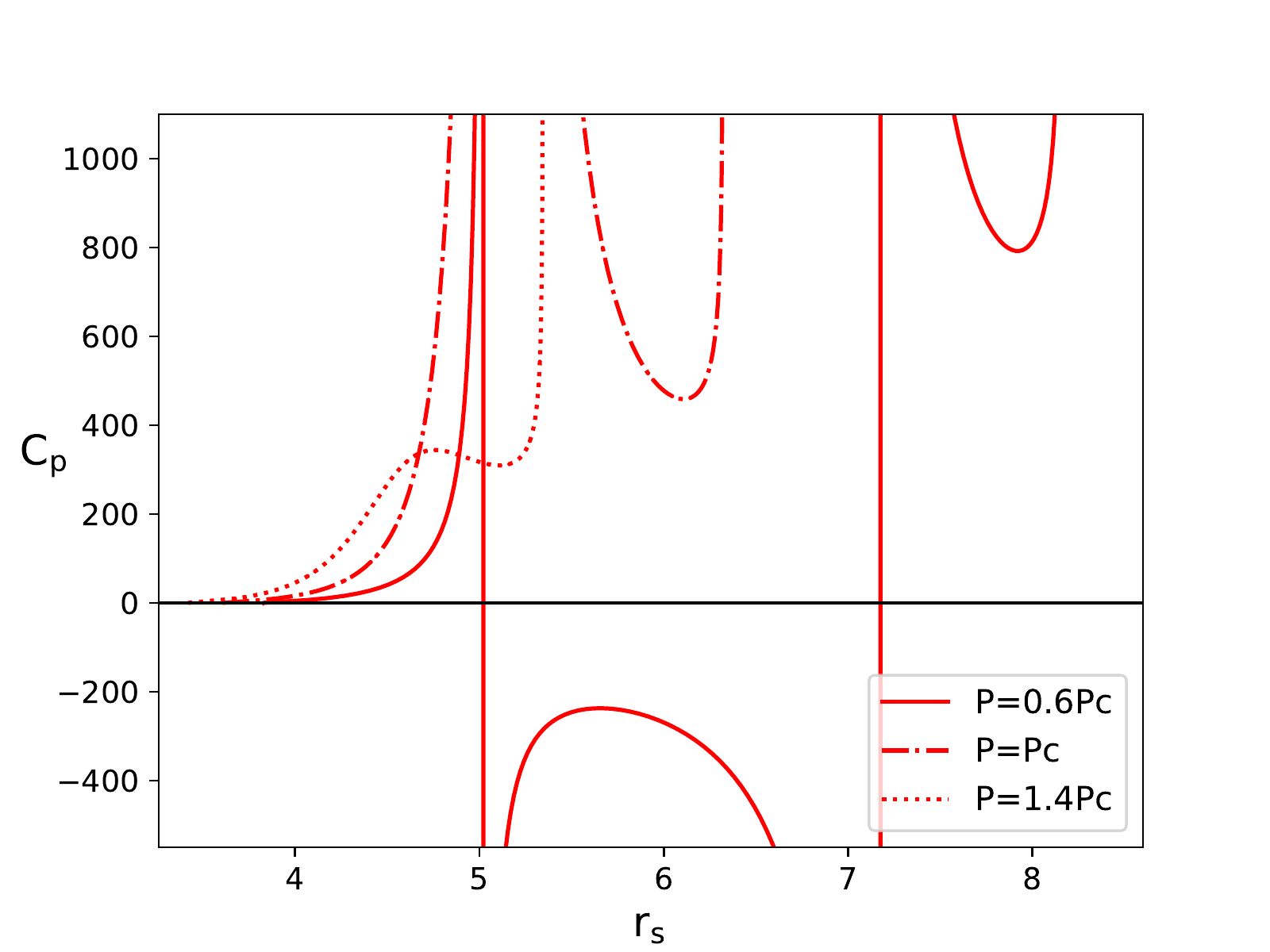}
			%\caption{fig2}
		\end{minipage}
	}%	
	
	\caption{The heat capacity $C_P$ versus $r_h$ (\ref{fig31}, ~\ref{fig32}, ~\ref{fig33}) and $r_{s}$ (\ref{fig34}, ~\ref{fig35}, ~\ref{fig36}) for the magnetic charge $g = 0.3,~0.6,~0.9$. The solid, dashed and dotted lines correspond to $P=0.6P_{c}$, $P=P_{c}$ and $P=1.4P_{c}$, respectively.}
    \label{fig3}
\end{figure}

\section{Thermal profile of the Hayward-AdS BH}
\label{sec:4}
In order to show the phase structure of the Hayward-AdS BH more intuitively, we calculate the boundary curve of the shadow of the Hayward-AdS BH. For such a spherically symmetric BH, the shape of the BH shadow is circular for an arbitrary observer~\cite{Chang:2020lmg}. Therefore, one can use a thermal profile in a two-dimensional plane to study the relation between the thermodynamic PT and shadow of the Hayward-AdS BH. In this case, for the stereographic projection of the Hayward-AdS BH located at in the celestial coordinate onto the two-dimensional plane, the boundary curve of the shadow is given as follows\cite{Eiroa:2017uuq},  
\begin{equation}
	\label{3-2-11}
	x=\lim _{r \rightarrow \infty}\left(-r^2 \sin \theta_0 \frac{d \phi}{d r}\right)_{\theta_0 \rightarrow \frac{\pi}{2}}, \quad y=\lim _{r \rightarrow \infty}\left(r^2 \frac{d \theta}{d r}\right)_{\theta_0 \rightarrow \frac{\pi}{2}},
\end{equation}
where $x$ and $y$ are the Cartesian coordinates with the inclination angle $\theta_0 \rightarrow \frac{\pi}{2}$.

The shadow contours of the Hayward-AdS BH for different parameters are shown in Fig.~\ref{fig4}, where we have set $M=60$ and $f(r_{o}=100)=1$ for the static observer. The results show that the shadow radius consistently increases with the pressure. The solid lines denote $P<P_{c}$, which correspond to the largest shadow radius region. As we clarified earlier, in this case, there exits a vdW-like PT between the small BH and large BH. The dashed lines represent $P=P_{c}$, which is a critical point because there is an unstable PT. The dotted lines correspond to $P>P_{c}$ and the shadow represents a supercritical phase, for which no PT occurs. Moreover, for the same pressure, comparing the radii of the shadow contours with different magnetic charges, one can find that the shadow radius of the Hayward-AdS BH increases with the magnetic charge $g$.
\begin{figure}[]
	\centering

	\subfigure[$g=0.3$\label{fig41}]{
		\begin{minipage}[t]{0.32\linewidth}
			\centering
			\includegraphics[width=2.00in]{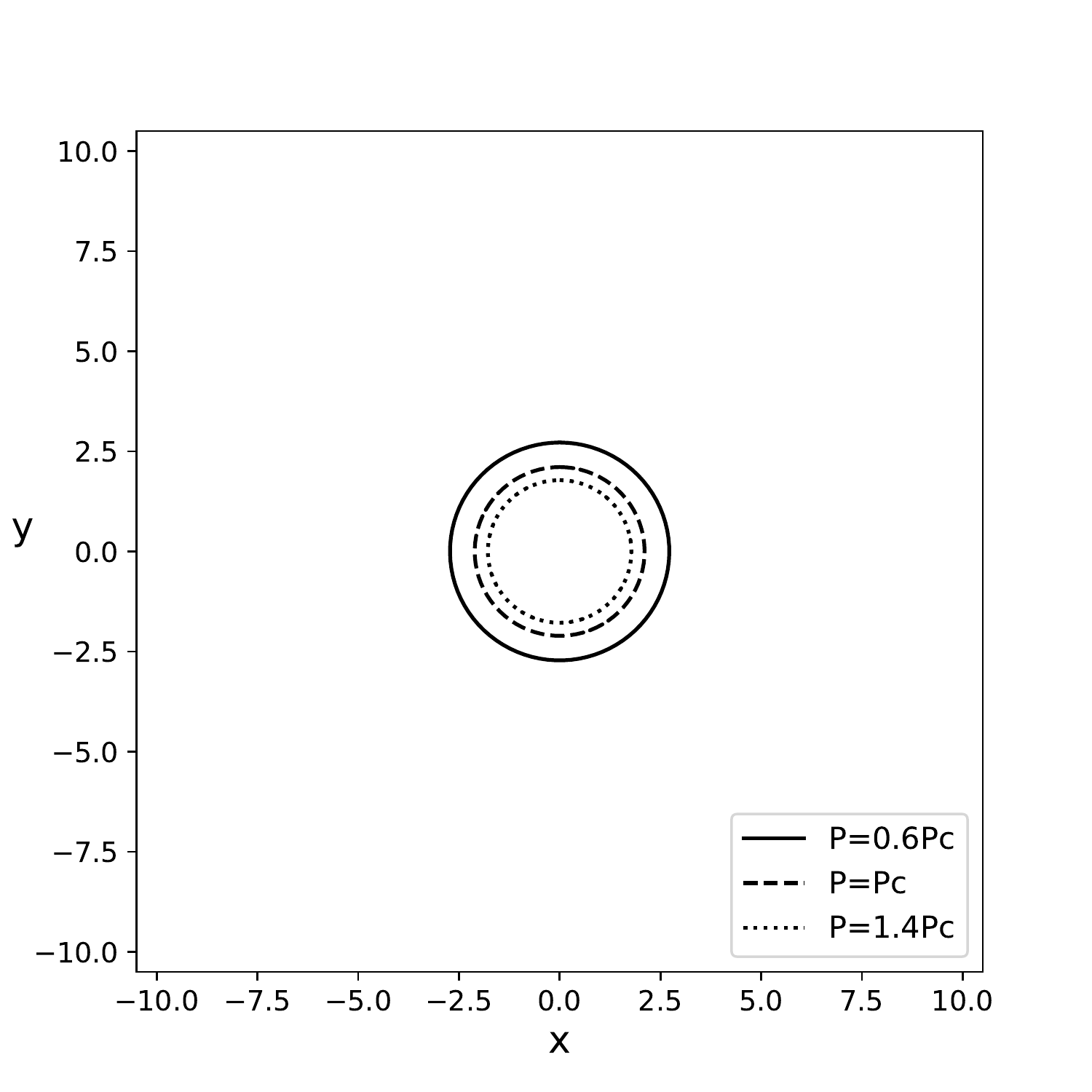}
			%\caption{fig1}
		\end{minipage}%
	}%
	\subfigure[$g=0.6$\label{fig42}]{
		\begin{minipage}[t]{0.32\linewidth}
			\centering
			\includegraphics[width=2.00in]{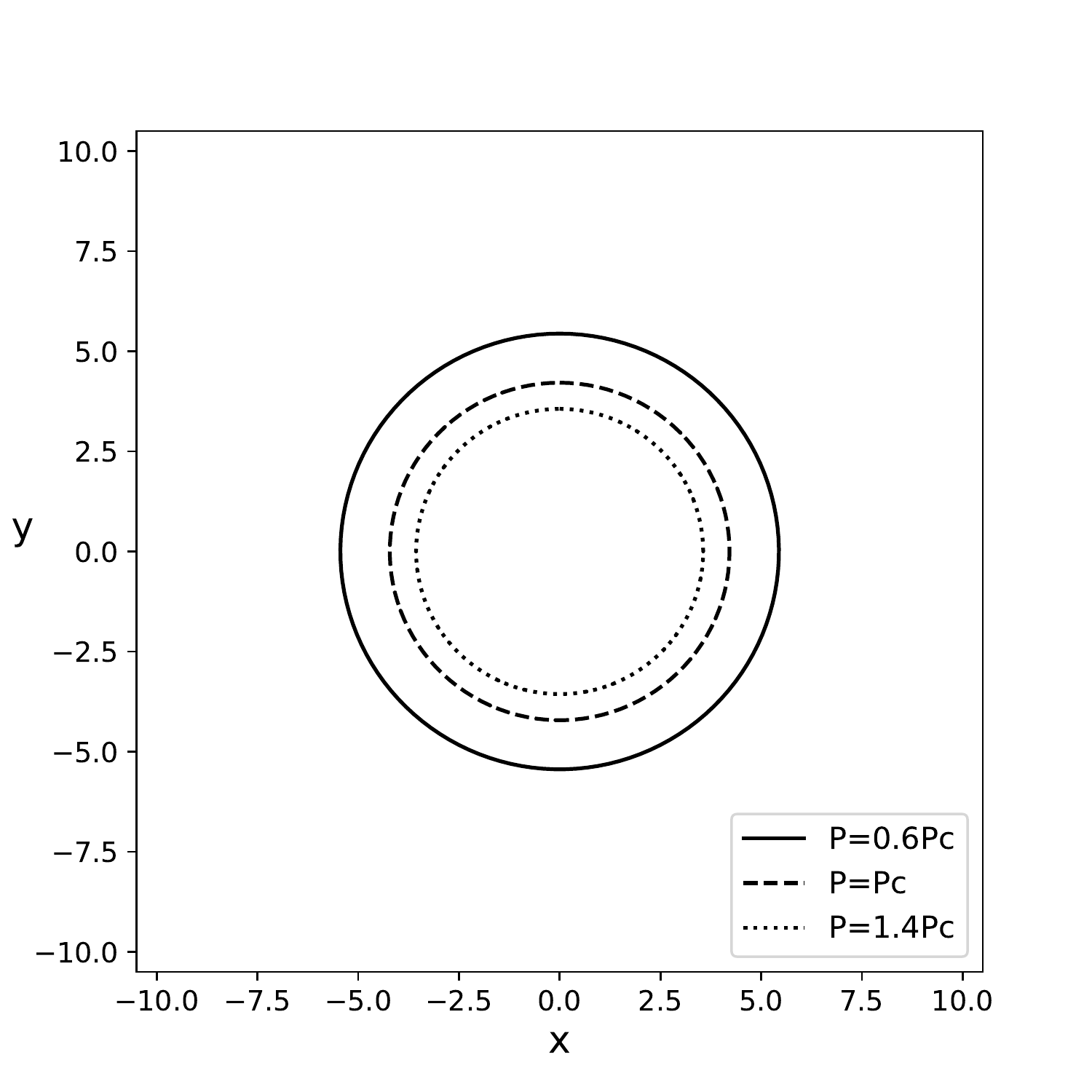}
			%\caption{fig2}
		\end{minipage}%
	}%
	\subfigure[$g=0.9$\label{fig43}]{
		\begin{minipage}[t]{0.32\linewidth}
			\centering
			\includegraphics[width=2.00in]{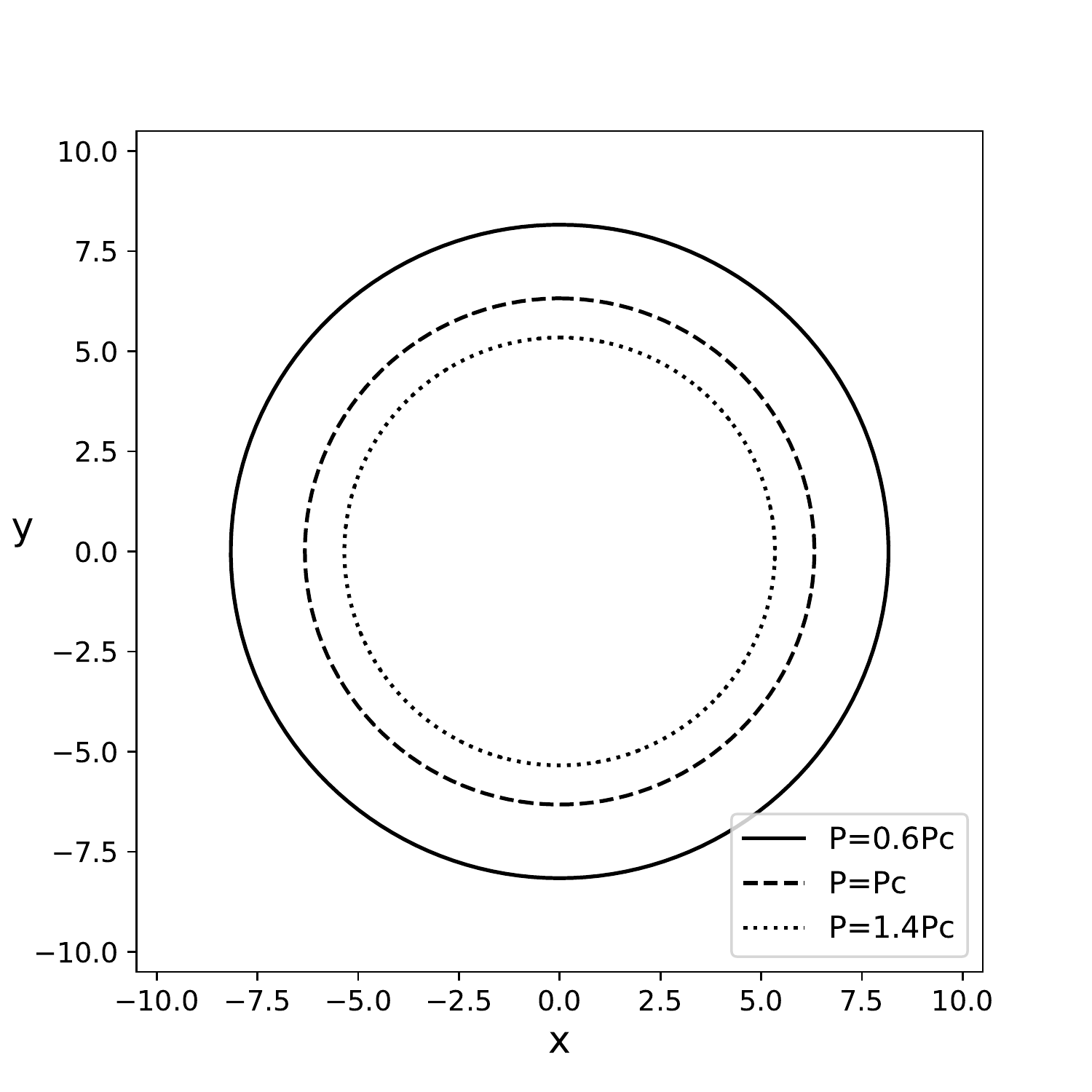}
			%\caption{fig2}
		\end{minipage}
	}%	
	\caption{The boundary curve of the shadow of the Hayward-AdS BH for the magnetic charge $g = 0.3,~0.6,~0.9$.  The solid, dashed and dotted lines correspond to $P=0.6P_{c}$, $P=P_{c}$ and $P=1.4P_{c}$, respectively. We have set $M=60$ and $r_{o}=100$.}
	\label{fig4}
\end{figure}

To analyze further, from multiple aspects, the relation between the phase structure and the shadow of the Hayward-AdS BH, one can study the superposition of the temperature diagram (see Fig.~\ref{fig2}) and the shadow cast diagram (see Fig.~\ref{fig4}), which is called the thermal profile of the Hayward-AdS BH (see Fig.~\ref{fig5}). In order to do that, we need to fix the pressure of the Hayward-AdS BH
and then the shadow contour can be considered as a function of temperature for a static observer located at position $r_o = 100$. Fig.~\ref{fig5} illustrate the thermal profiles of the Hayward-AdS BH for three sets of magnetic charges and three sets of pressures. Figs.~\ref{fig51}, \ref{fig54} and~\ref{fig57} represent $P<P_{c}$, which correspond to the solid lines in Fig.\ref{fig4}. It can be found that when the shadow radius is smaller than a certain value, the temperature increases from the center to the boundary of the shadow. Then, the temperature oscillates as the shadow radius increases, i.e., the N-type change trend~\cite{Guo:2022yjc}. Figs.~\ref{fig52}, \ref{fig55} and~\ref{fig58} denote $P=P_{c}$, which correspond to the dashed lines in Fig.\ref{fig4}. It exits a critical region of thermodynamic instability, where the temperature remains constant. Figs.~\ref{fig53}, \ref{fig56} and~\ref{fig59} represent $P>P_{c}$, which correspond to the dotted lines in Fig.\ref{fig4}. One can find that the temperature keeps increasing from the center to the boundary of the shadow, which consists of the result of the dotted lines in Fig.~\ref{fig2}. In addition, comparing these diagrams, one can find that when the magnetic charge $g$ is fixed, the temperature increases with the pressure while the region of the thermal profile decreases with the pressure. 

Finally, for the Hayward-AdS BH, we study the region where the N-type change trend occurs. For the case of $P<P_{c}$, the thermal profiles of the Hayward-AdS BH with different magnetic charges from $r_{s1}$ to $r_{s2}$ are shown in Fig.~\ref{fig6}. It can be found that the temperature increases first, then decreases, and finally increases again from the center to the boundary of the shadow, which is consistent with the result of the solid lines in Fig.~\ref{fig2}. The results imply that when $P<P_{c}$, the thermal profile can be used to investigate the phase structure of the Hayward-AdS BH.

\begin{figure}[h]
	\centering

	\subfigure[$P=0.6P_{ c}$ and $g=0.3$\label{fig51}]{
		\begin{minipage}[t]{0.25\linewidth}
			\centering
			\includegraphics[width=1.8in]{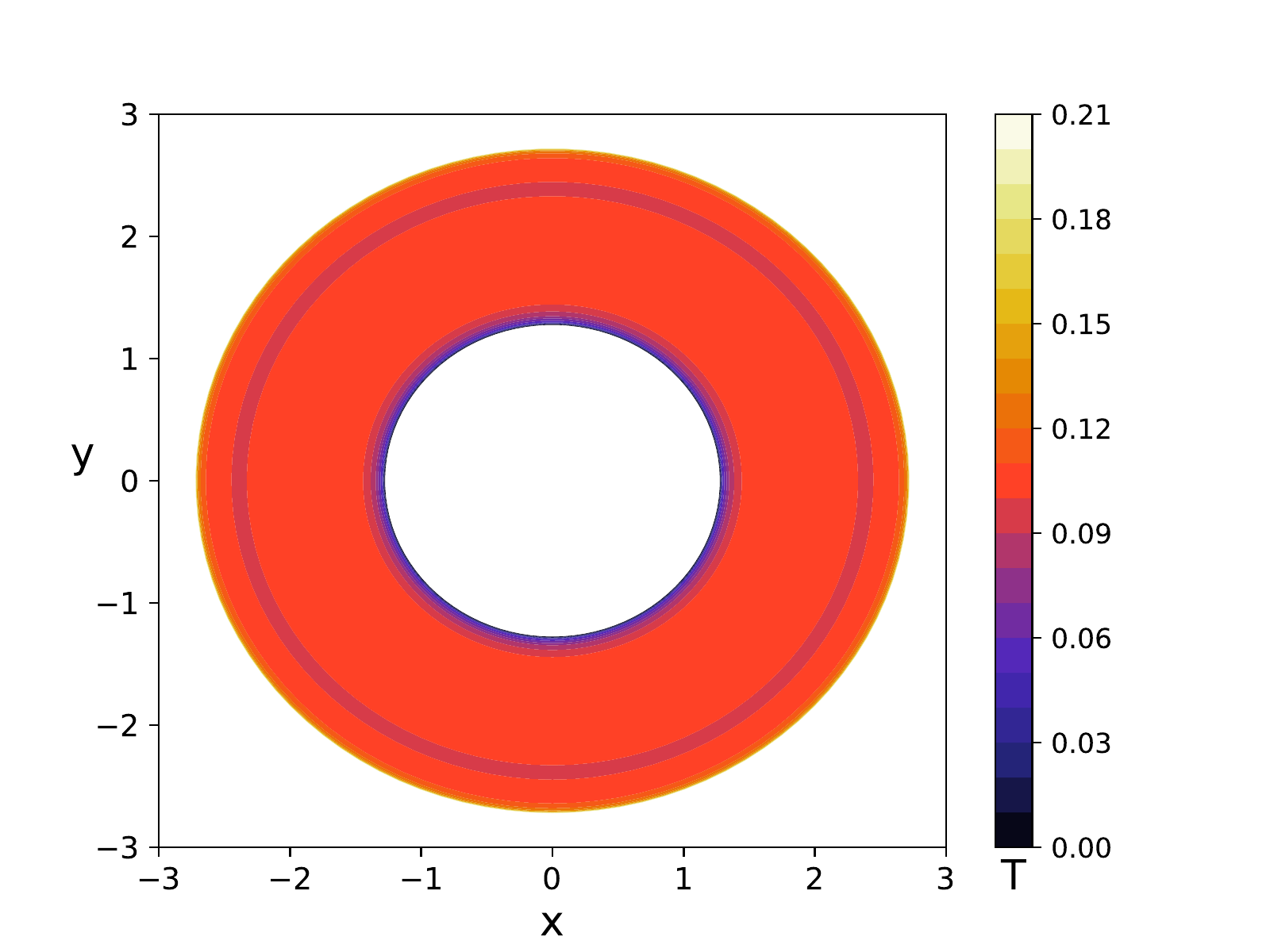}
			%\caption{fig1}
		\end{minipage}%
	}%
	\subfigure[$P=P_{ c}$ and $g=0.3$\label{fig52}]{
		\begin{minipage}[t]{0.25\linewidth}
			\centering
			\includegraphics[width=1.8in]{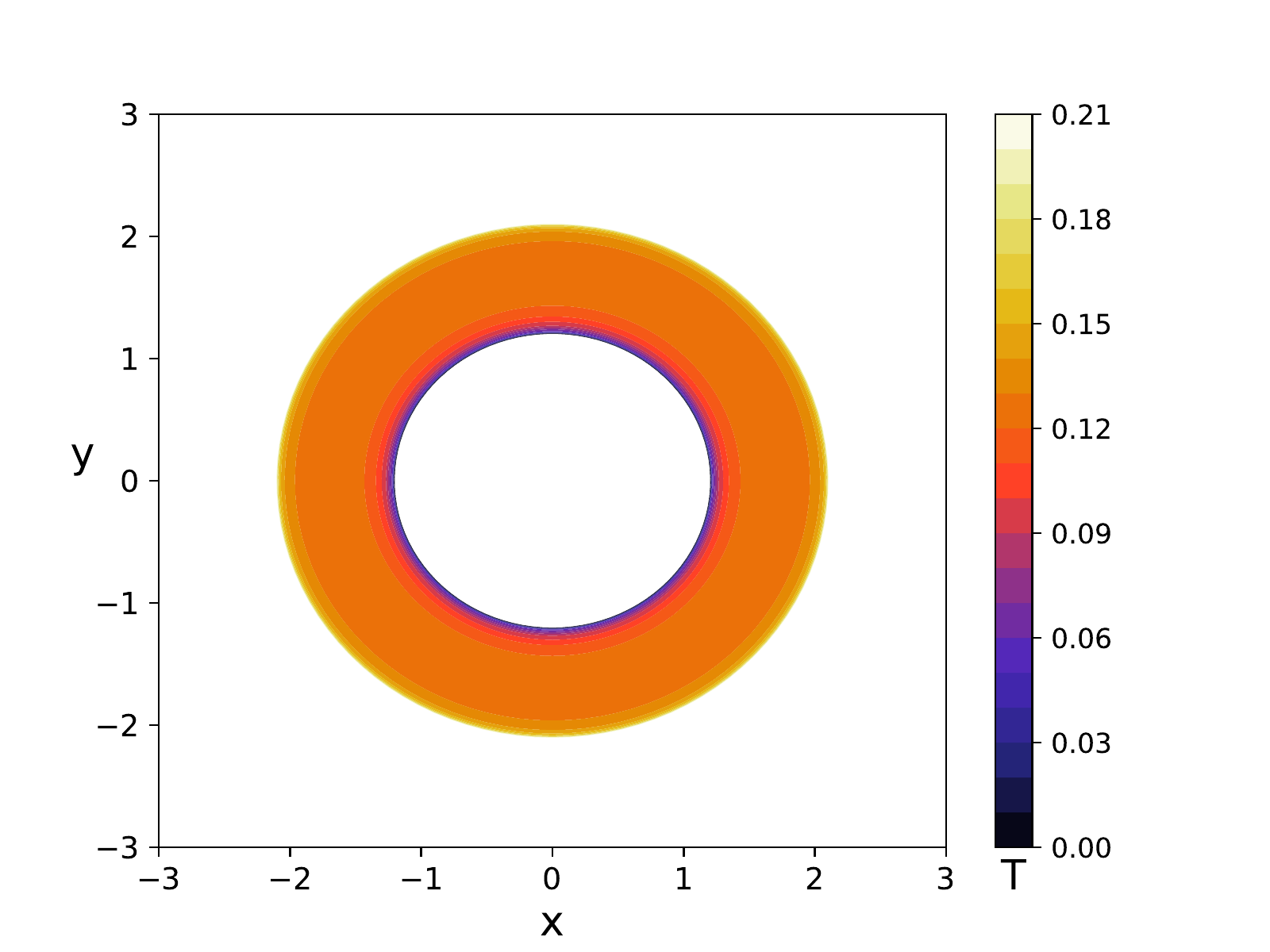}
			%\caption{fig2}
		\end{minipage}%
	}%
	\subfigure[$P=1.4P_{ c}$ and $g=0.3$\label{fig53}]{
		\begin{minipage}[t]{0.25\linewidth}
			\centering
			\includegraphics[width=1.8in]{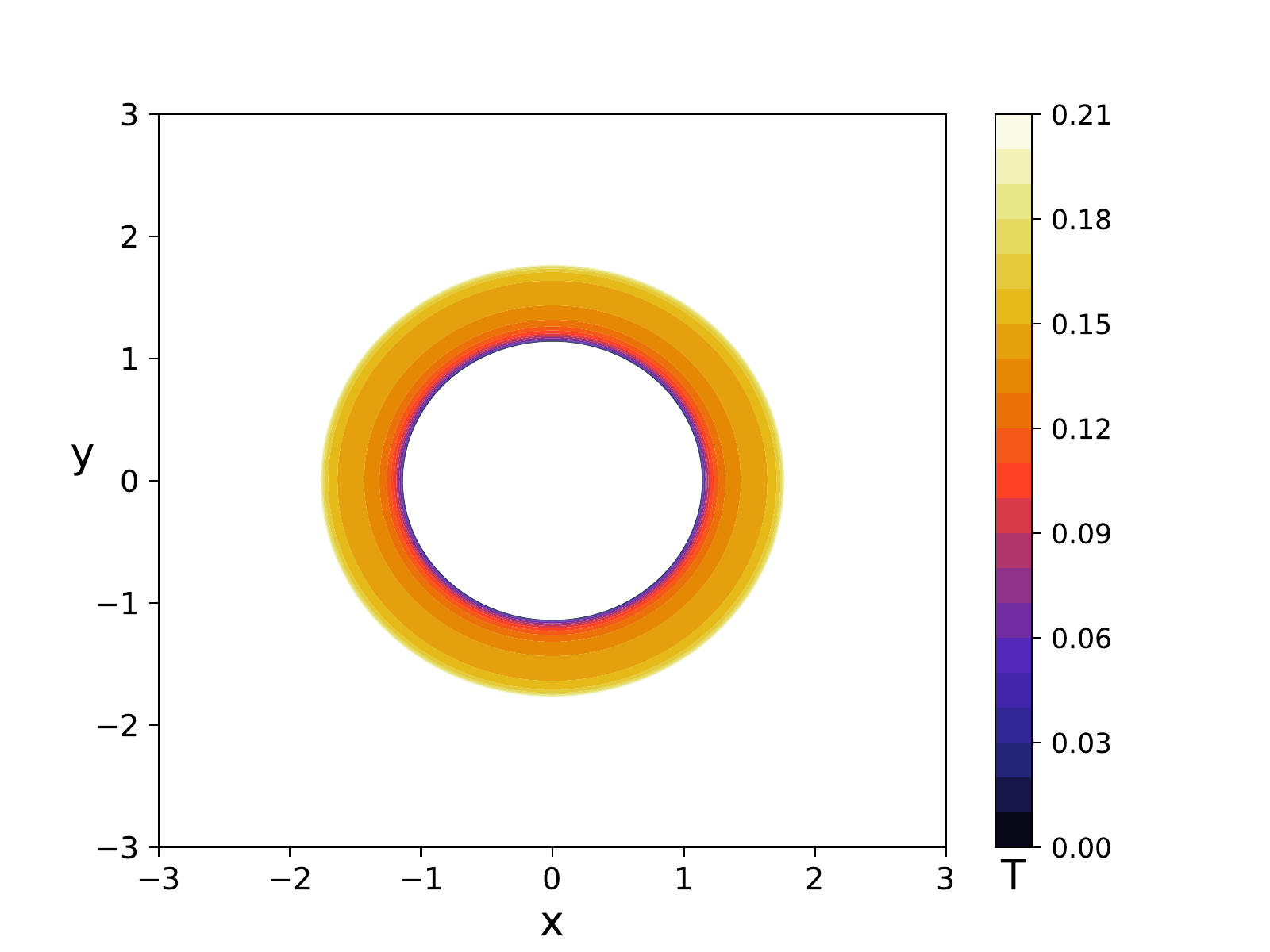}
			%\caption{fig2}
		\end{minipage}
	}%
	\quad
	\subfigure[$P=0.6P_{ c}$ and $g=0.6$\label{fig54}]{
		\begin{minipage}[t]{0.25\linewidth}
			\centering
			\includegraphics[width=1.8in]{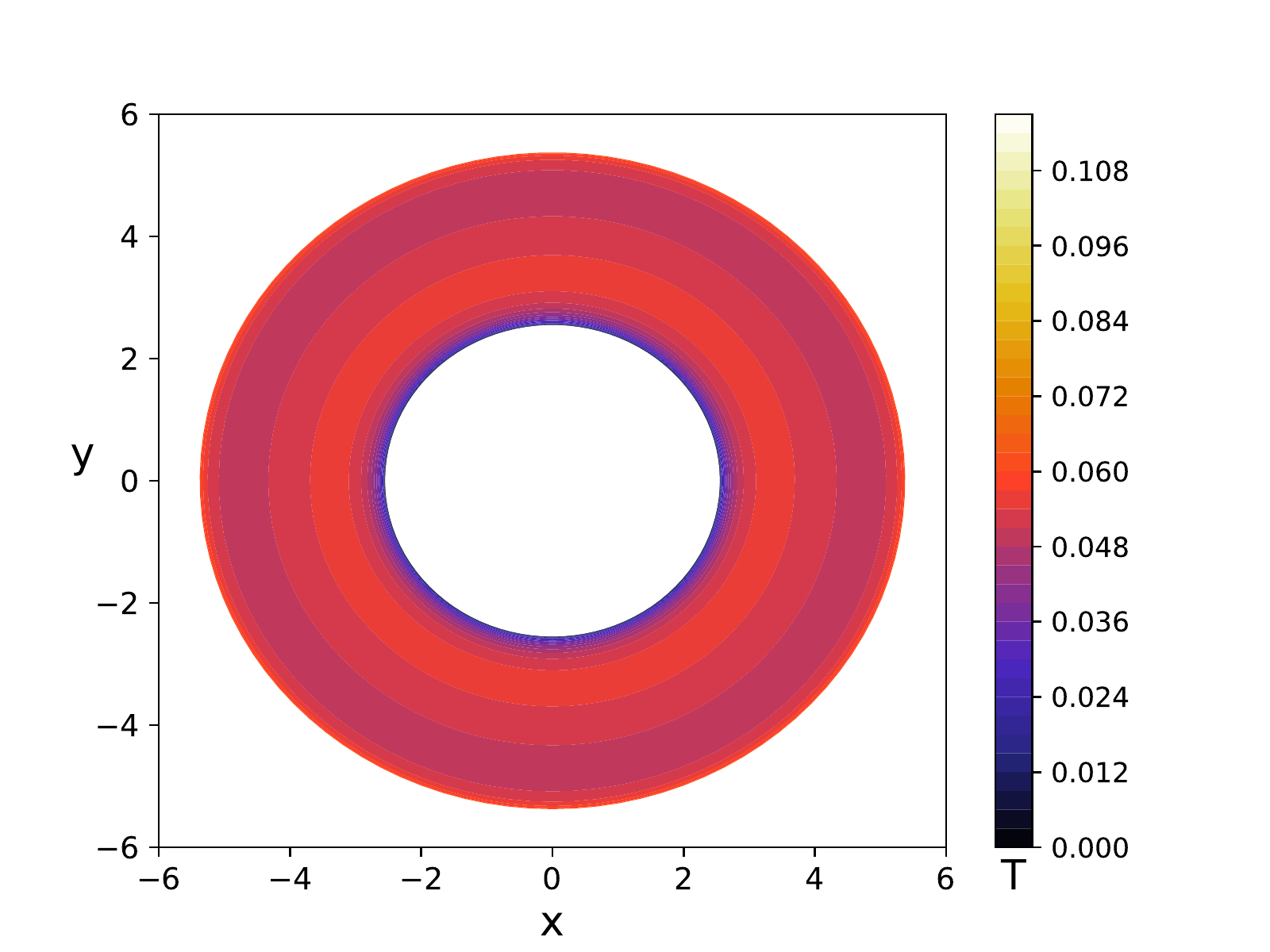}
			%\caption{fig2}
		\end{minipage}
	}%
	\subfigure[$P=P_{ c}$ and $g=0.6$\label{fig55}]{
		\begin{minipage}[t]{0.25\linewidth}
			\centering
			\includegraphics[width=1.8in]{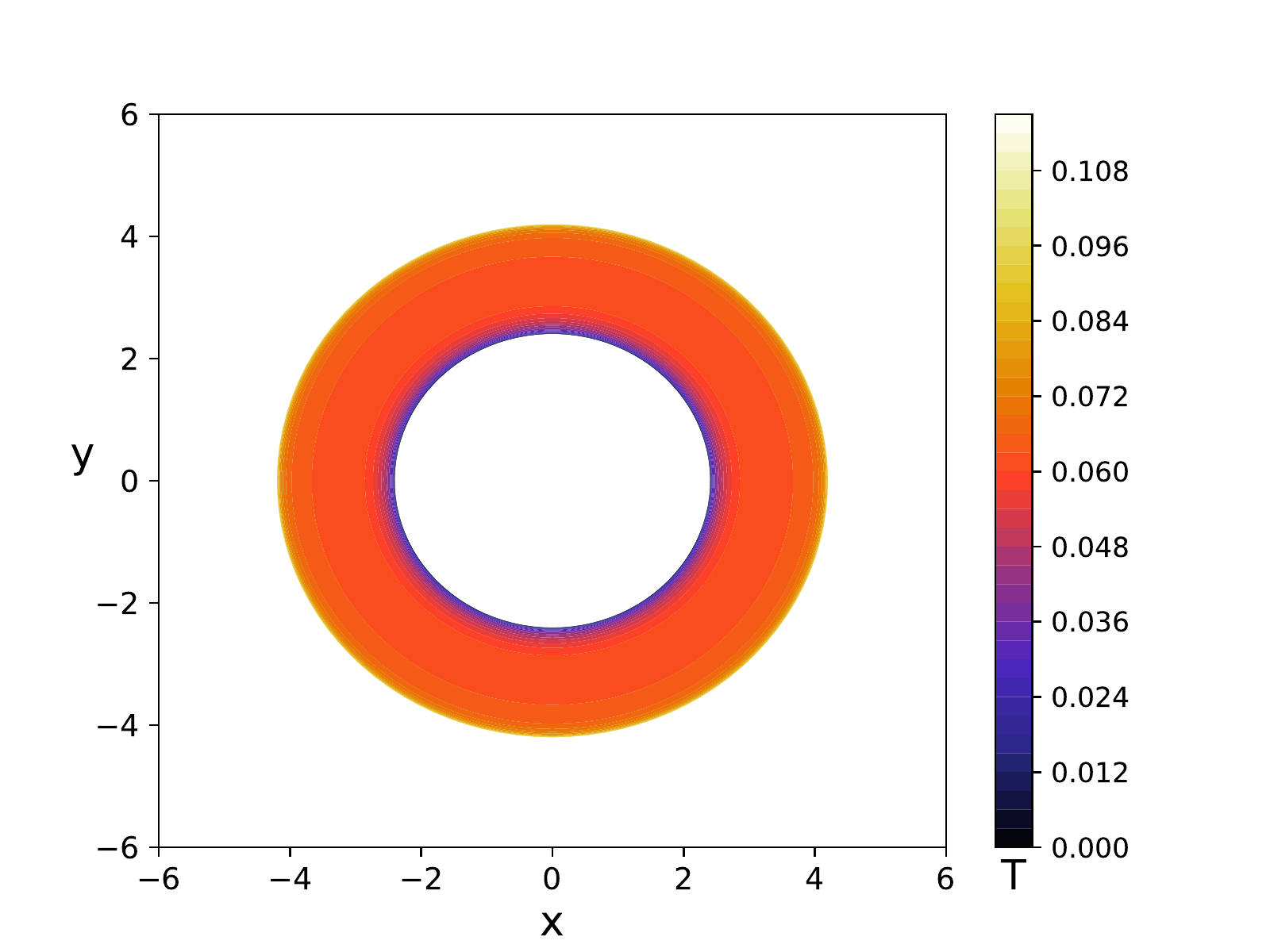}
			%\caption{fig2}
		\end{minipage}
	}%
	\subfigure[$P=1.4P_{ c}$ and $g=0.6$\label{fig56}]{
		\begin{minipage}[t]{0.25\linewidth}
			\centering
			\includegraphics[width=1.8in]{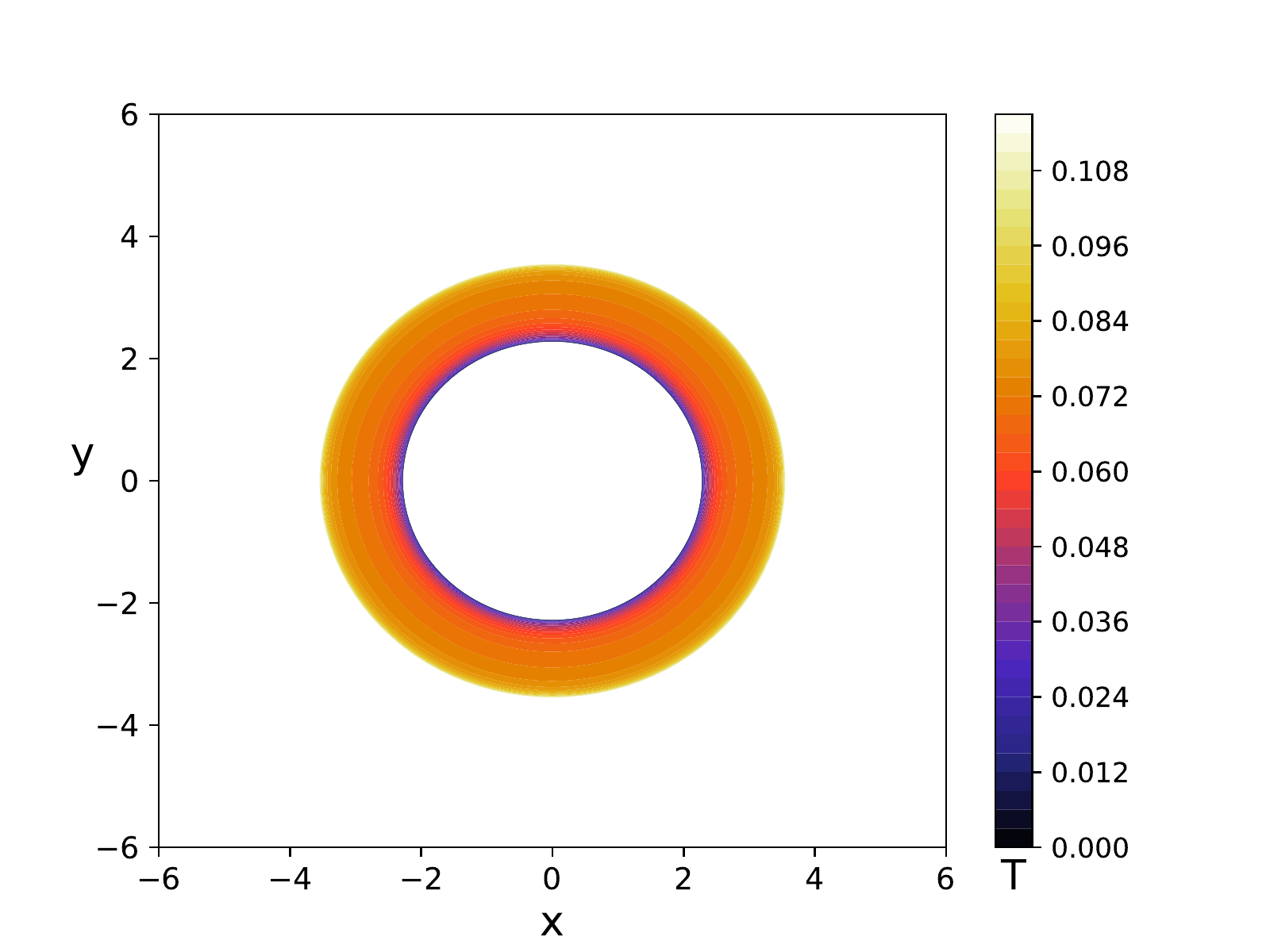}
			%\caption{fig2}
		\end{minipage}
	}%	
    \quad
    \subfigure[$P=0.6P_{ c}$ and $g=0.9$\label{fig57}]{
	\begin{minipage}[t]{0.25\linewidth}
		\centering
		\includegraphics[width=1.8in]{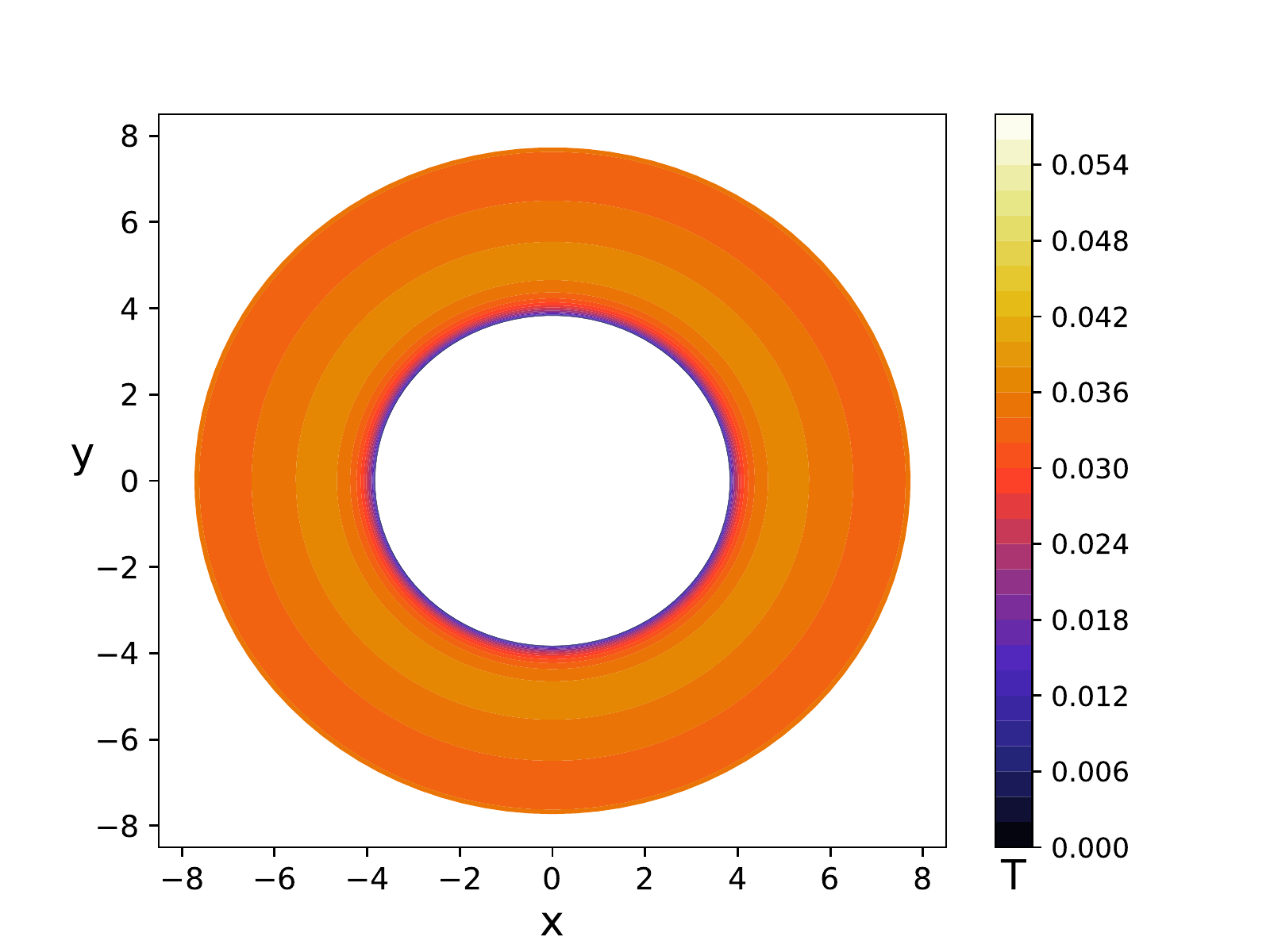}
		%\caption{fig2}
	\end{minipage}
    }%
    \subfigure[$P=P_{ c}$ and $g=0.9$\label{fig58}]{
	\begin{minipage}[t]{0.25\linewidth}
		\centering
		\includegraphics[width=1.8in]{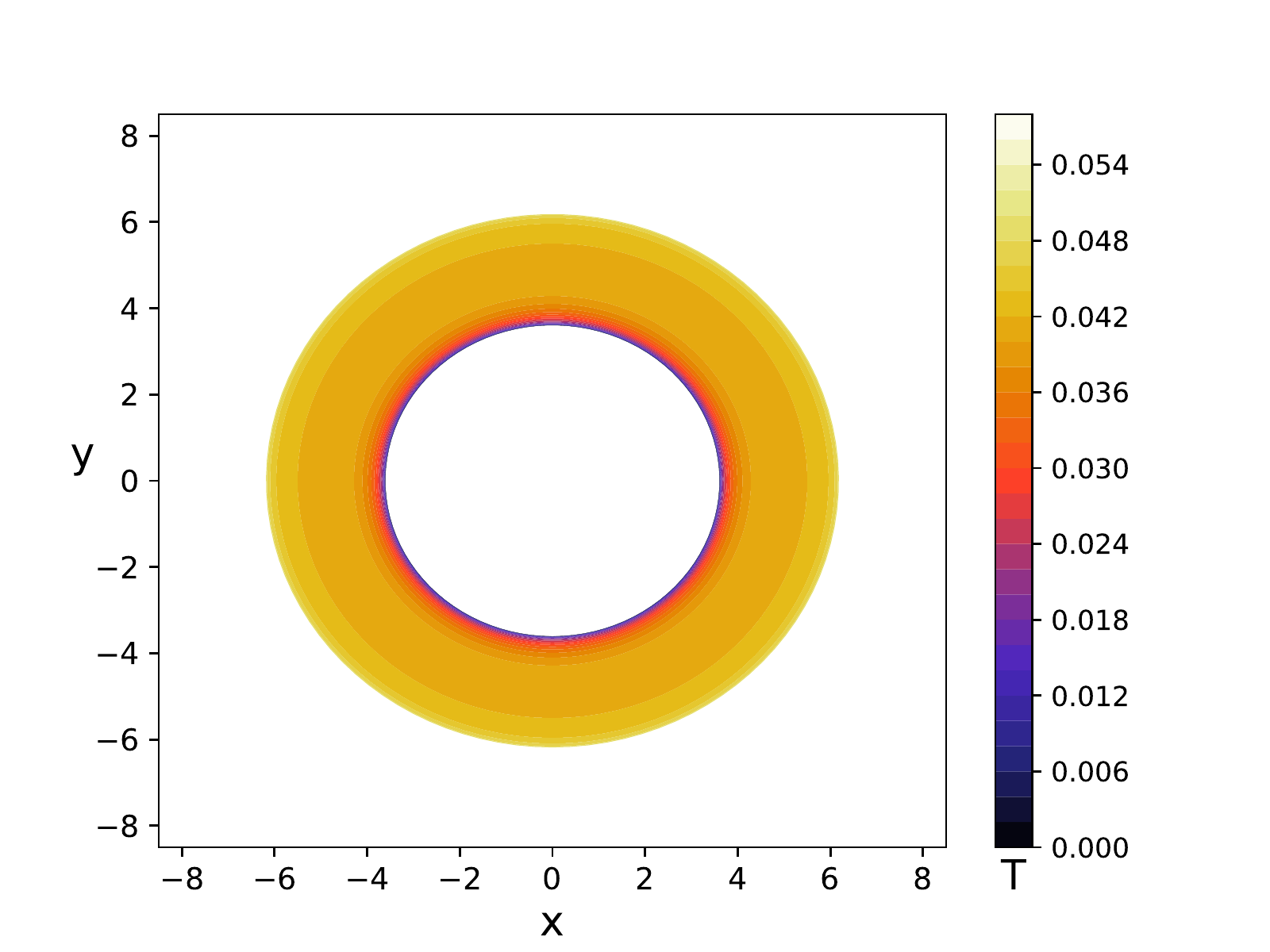}
		%\caption{fig2}
	\end{minipage}
    }%
    \subfigure[$P=1.4P_{ c}$ and $g=0.9$\label{fig59}]{
	\begin{minipage}[t]{0.25\linewidth}
		\centering
		\includegraphics[width=1.8in]{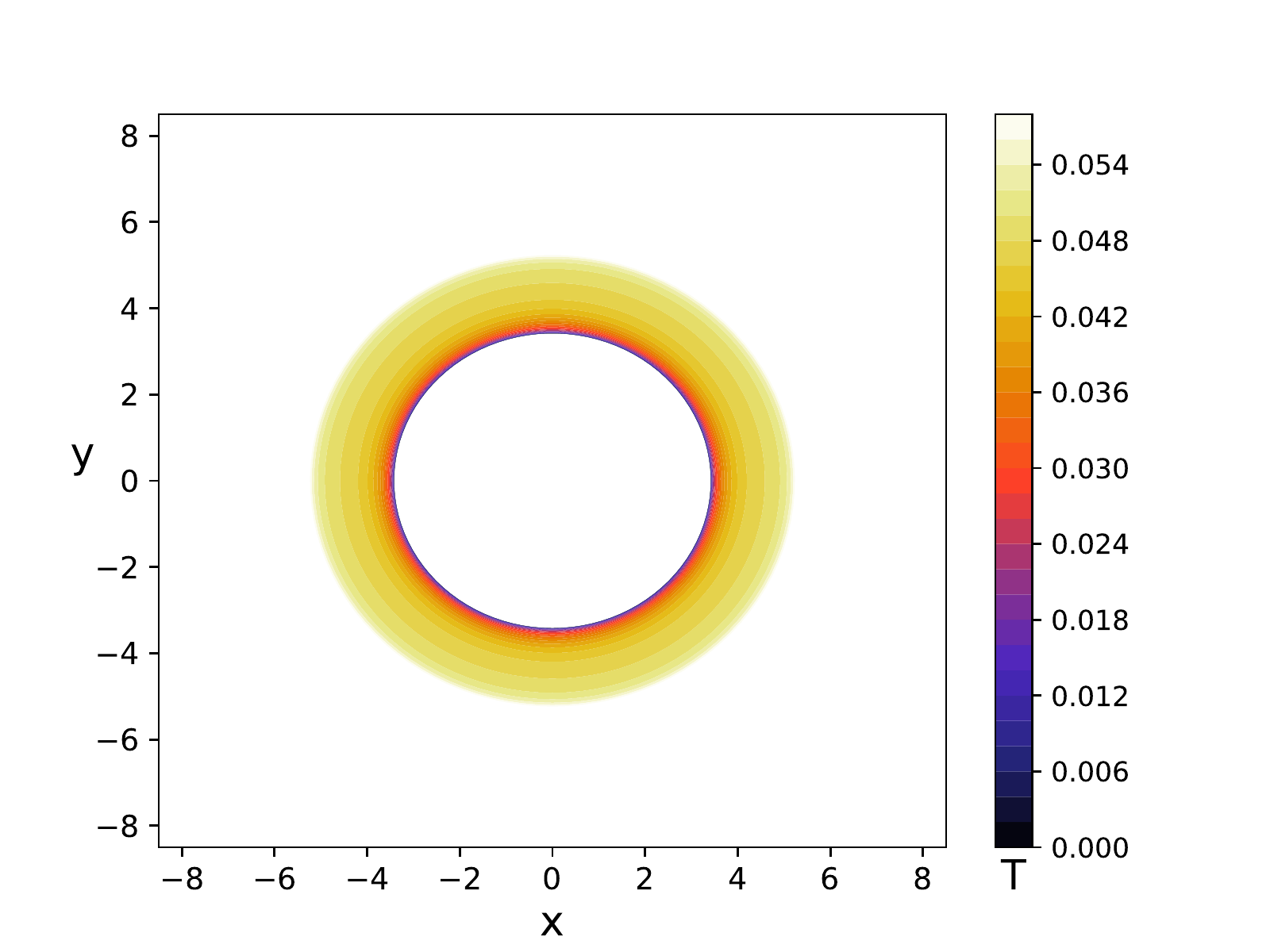}
		%\caption{fig2}
	\end{minipage}
}%	
	
	\caption{Thermal profiles of the Hayward-AdS BH for the magnetic charge $g = 0.3,~0.6,~0.9$ and the pressure $P=0.6P_c$, $P_c$, $1.4P_c$. We have set $M=60$ and $r_{o}=100$.}
	\label{fig5}
\end{figure}

\begin{figure}[h]
	\centering

	\subfigure[$P=0.6P_{ c}$ and $g=0.3$\label{fig61}]{
		\begin{minipage}[t]{0.32\linewidth}
			\centering
			\includegraphics[width=2.00in]{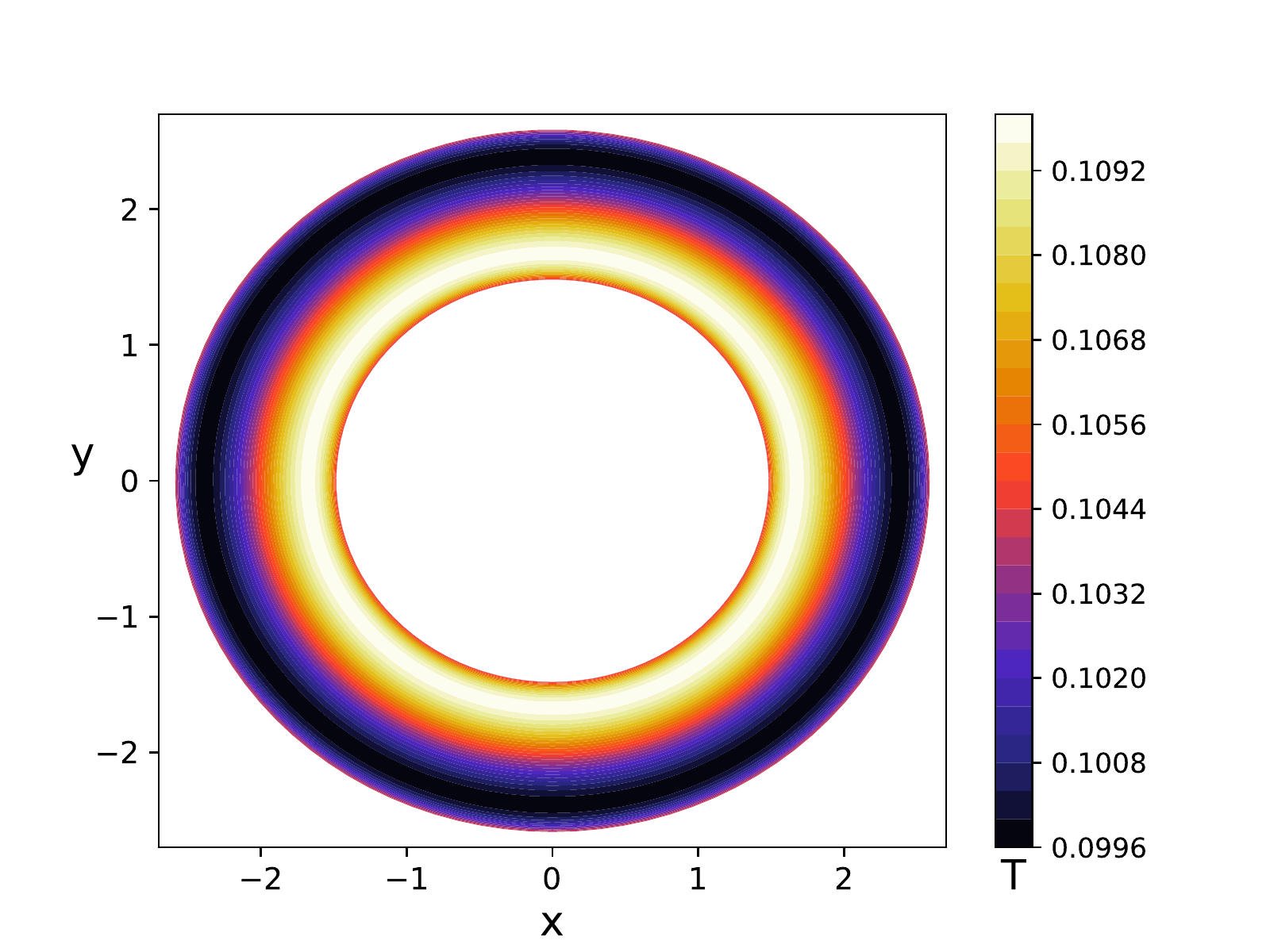}
			%\caption{fig1}
		\end{minipage}%
	}%
	\subfigure[$P=0.6P_{ c}$ and $g=0.6$\label{fig62}]{
		\begin{minipage}[t]{0.32\linewidth}
			\centering
			\includegraphics[width=2.00in]{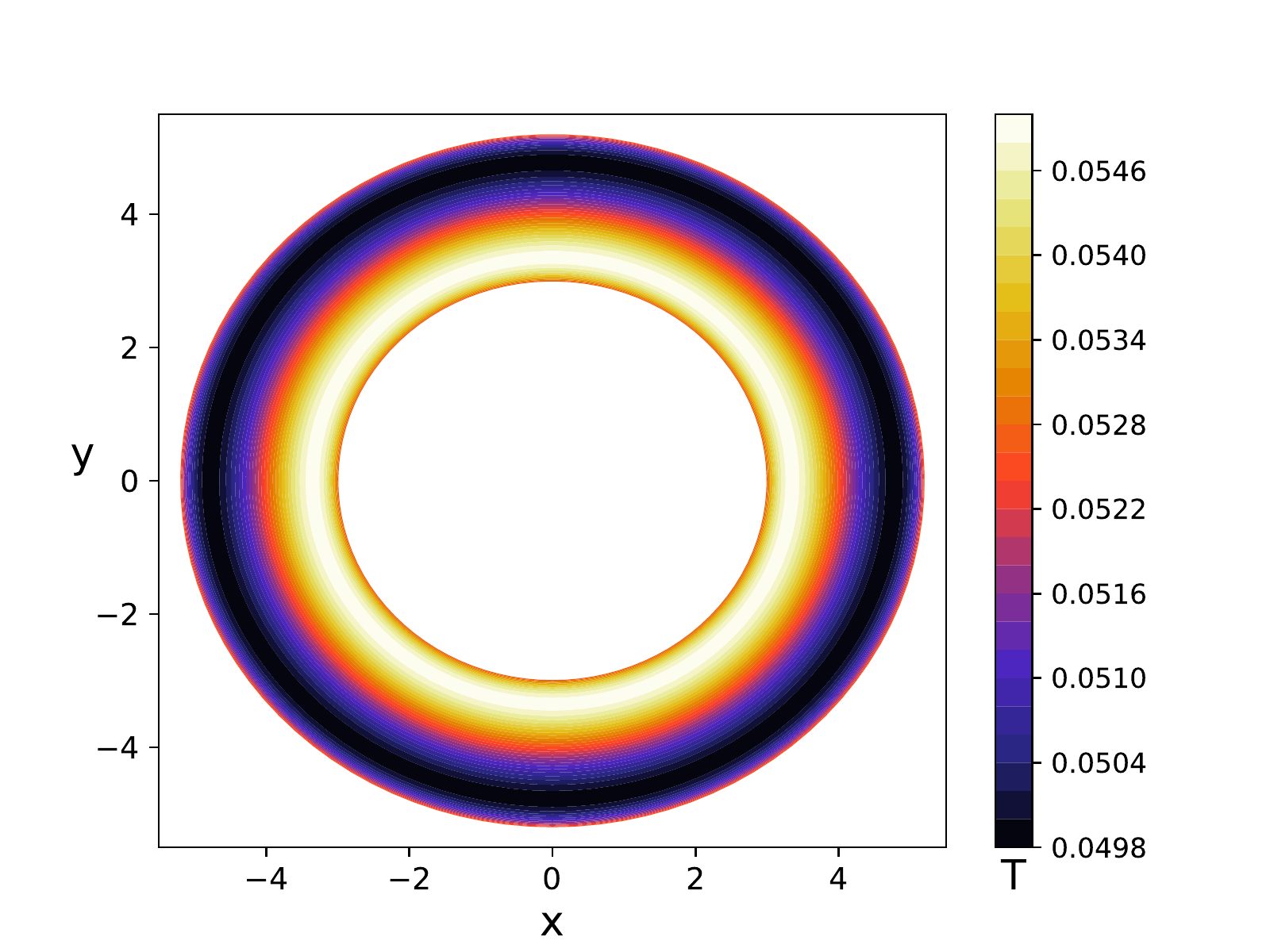}
			%\caption{fig2}
		\end{minipage}%
	}%
	\subfigure[$P=0.6P_{ c}$ and $g=0.9$\label{fig63}]{
		\begin{minipage}[t]{0.32\linewidth}
			\centering
			\includegraphics[width=2.00in]{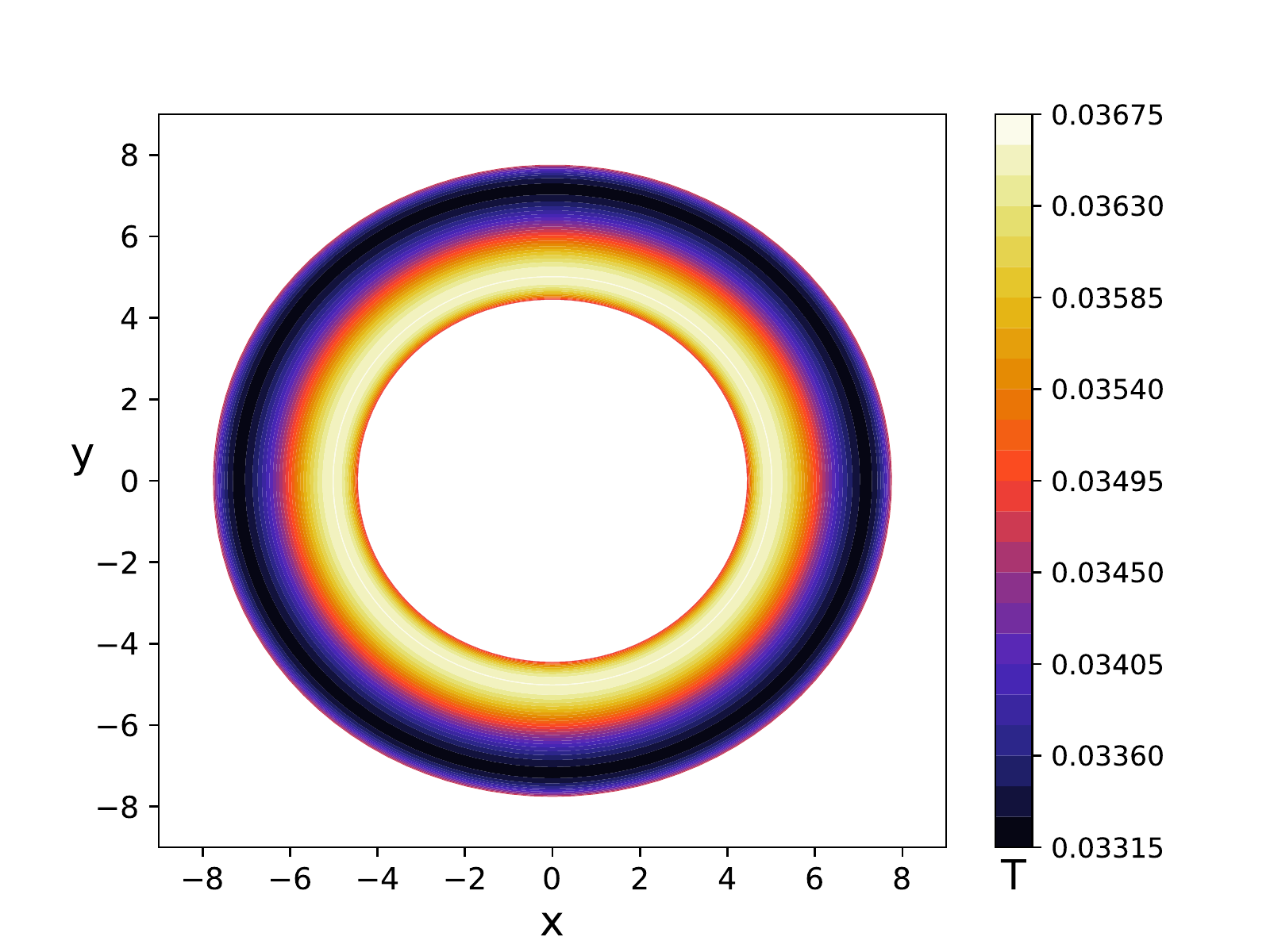}
			%\caption{fig2}
		\end{minipage}
	}%
	\quad
	\subfigure[$P=0.6P_{ c}$ and $g=0.3$\label{fig64}]{
		\begin{minipage}[t]{0.32\linewidth}
			\centering
			\includegraphics[width=2.00in]{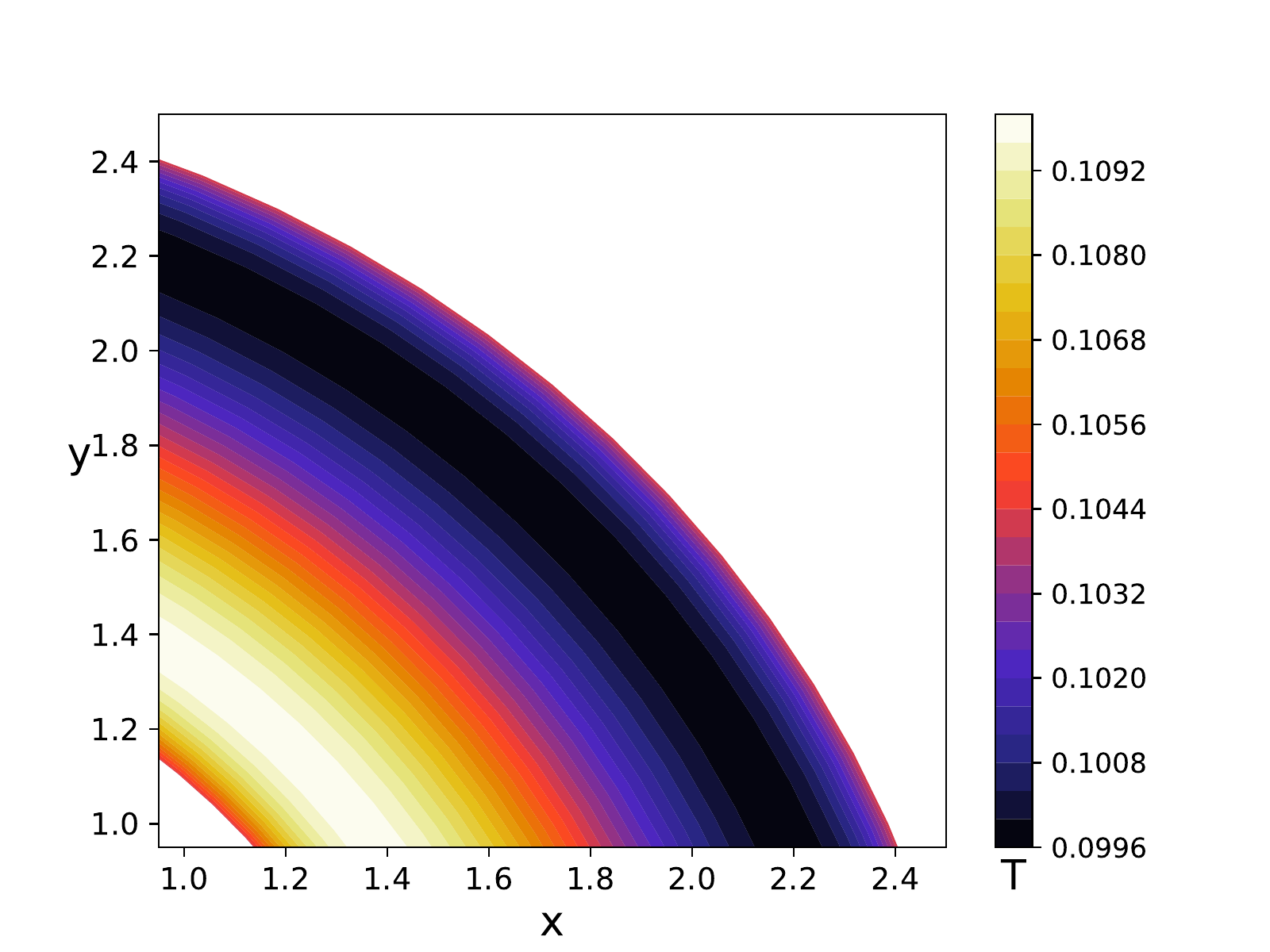}
			%\caption{fig2}
		\end{minipage}
	}%
	\subfigure[$P=0.6P_{ c}$ and $g=0.6$\label{fig65}]{
		\begin{minipage}[t]{0.32\linewidth}
			\centering
			\includegraphics[width=2.00in]{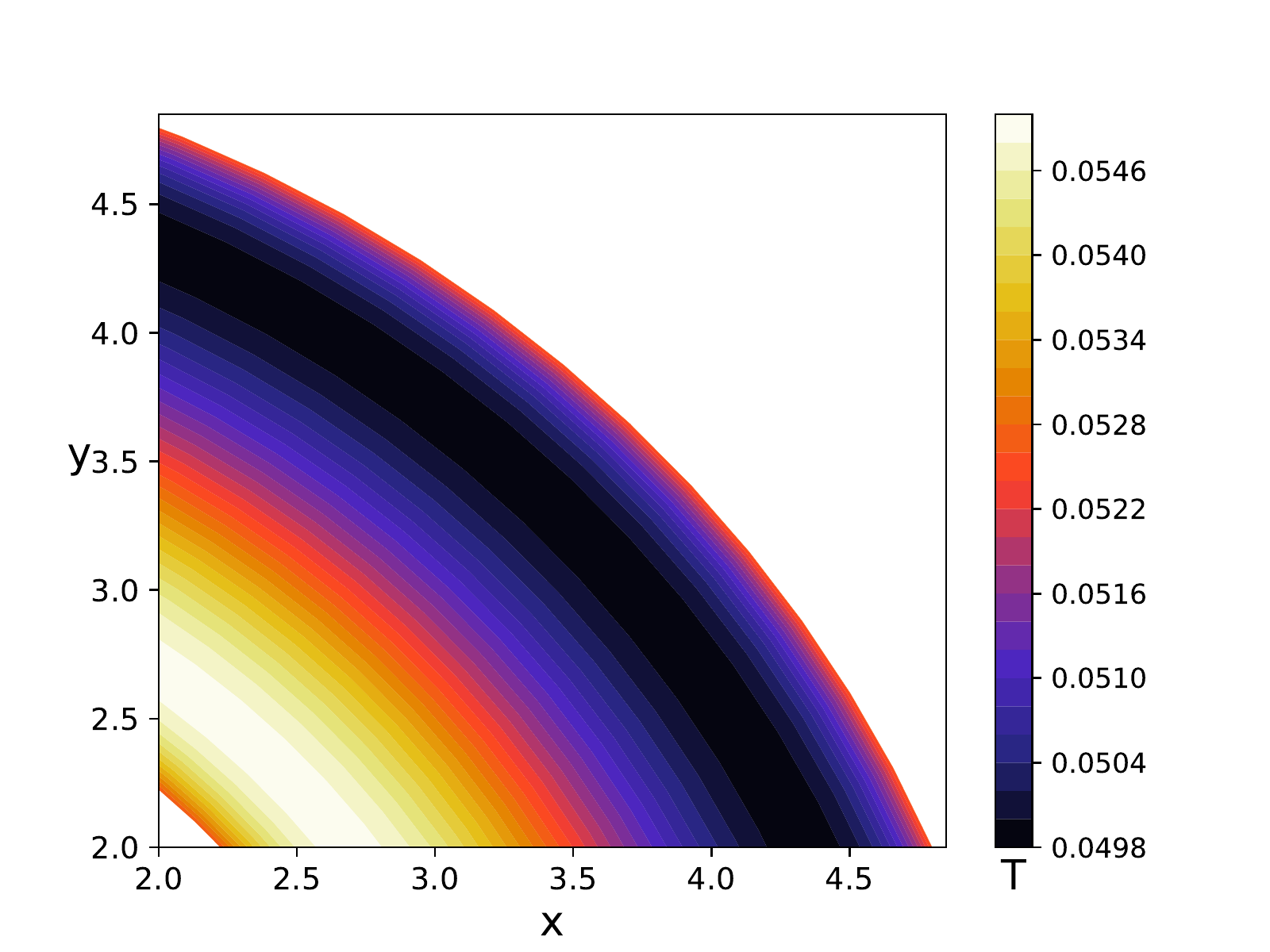}
			%\caption{fig2}
		\end{minipage}
	}%
	\subfigure[$P=0.6P_{ c}$ and $g=0.9$\label{fig66}]{
		\begin{minipage}[t]{0.32\linewidth}
			\centering
			\includegraphics[width=2.00in]{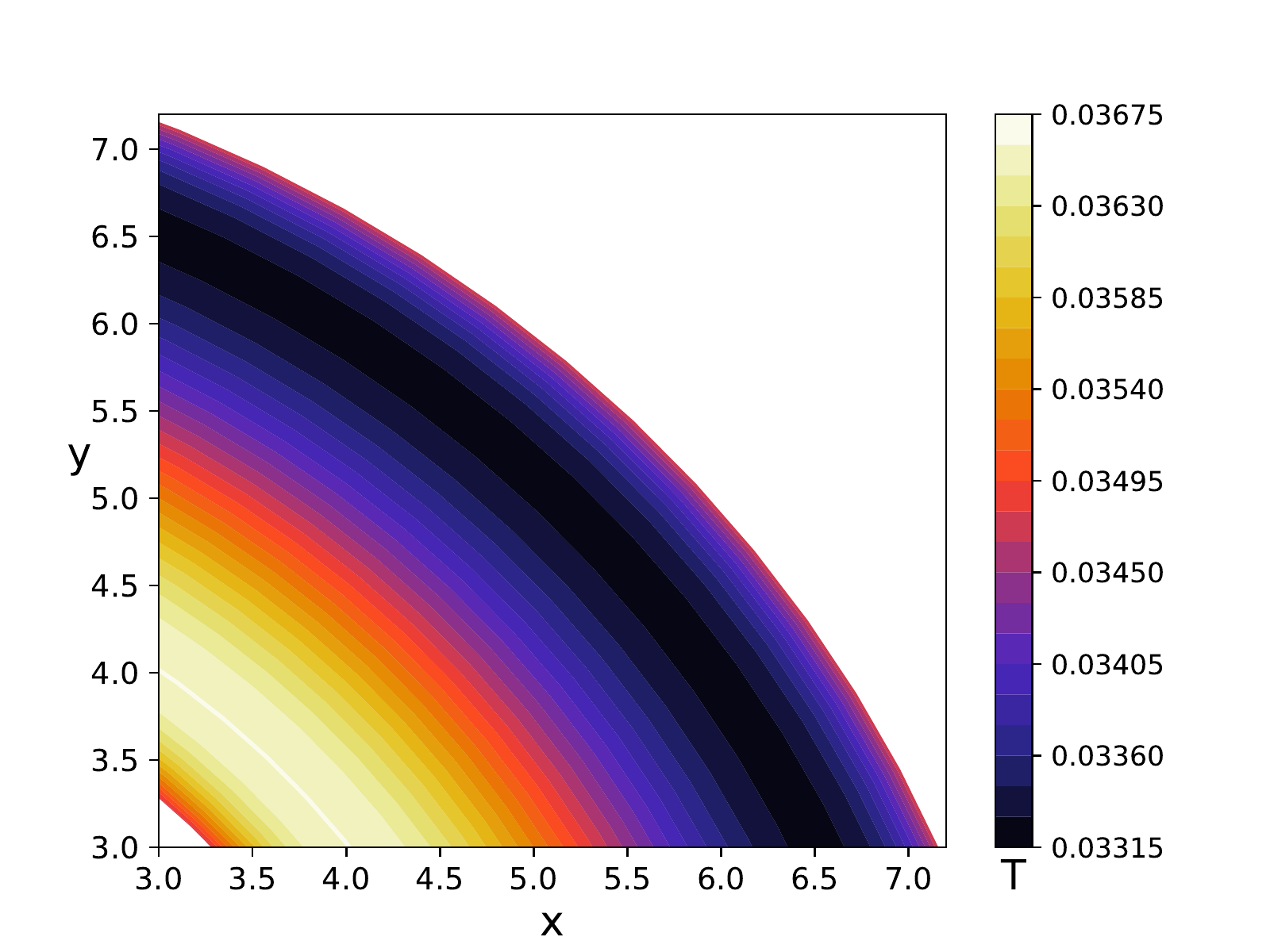}
			%\caption{fig2}
		\end{minipage}
	}%	
	
	\caption{Thermal profiles of the region from $r_{s1}$ to $r_{s2}$ for the magnetic charge $g = 0.3,~0.6,~0.9$ and the pressure $P=0.6P_c$. We have set $M=60$ and $r_{o}=100$.}
	\label{fig6}
\end{figure}

\section{Conclusions}\label{sec:5}
In this work, we investigate the relation between the phase structure of the Hayward-AdS BH and its shadow, which provides a new way to analyze the existence of the thermodynamic PT of the Hayward-AdS BH for different pressures. 

The expression for the shadow radius of the Hayward-AdS BH is derived from the Hamiltonian formalism. Then, we obtain the motion of the photon in the equatorial plane. According to the expression for the shadow radius, we plot the $r_{ s}-r_{ h}$ diagrams of the Hayward-AdS BH in Fig.~\ref{fig1}. Since the shadow radius is a monotonically increasing function of the horizon radius, the thermodynamics of the Hayward-AdS BH can be investigated by replacing the horizon radius $r_{h}$ with the shadow radius $r_{s}$. Moreover, the effect of the magnetic charge $g$ on the thermodynamics of the Hayward-AdS BH is studied. The results show that as the magnetic charge $g$ increases, the corresponding shadow radius $r_{ s}$ becomes larger. {In order to study the thermodynamics of the Hayward-AdS BH, the equation of state and the critical point of the PT process are calculated. As shown in Fig.~\ref{fig2}, for the pressure $P<P_{ c}$ the vdW-like PT occurs. The horizon radius regions $r_{h}<r_{h1}$, $r_{h1}<r_{h}<r_{h2}$ and $r_{h}>r_{h2}$ (the shadow radius regions $r_{s}<r_{s1}$, $r_{s1}<r_{s}<r_{s2}$ and $r_{s}>r_{s2}$) correspond to the small BH, intermediate BH and large BH, respectively. For the pressure $P=P_{ c}$, there exists an unstable PT. When the pressure $P>P_{ c}$, no PT occurs. Furthermore, from the whole Fig.~\ref{fig2}, one can find that for a fixed pressure $P$, as the magnetic charge $g$ increases, the shadow radius $r_{s}$ increases while the coexistence temperature $T_{co}$ decreases. 

To investigate the effect of the magnetic charge $g$ on the PT process of the Hayward-AdS BH from multiple aspects, we plot the shadow boundary curve in Fig.~\ref{fig4}. It is found that as the pressure increases, the shadow radius $r_{s}$ of the Hayward-AdS BH increases. For a fixed pressure $P$, the shadow radius $r_{s}$ increases with the magnetic charge $g$. Moreover, combining the temperature diagram and the shadow cast diagram, we can plot the thermal profiles of the Hayward-AdS BH (see Fig.~\ref{fig5}). For the pressure $P<P_{c}$, the N-type change trend of the BH temperature appears. For the case of $P=P_{c}$, there exits a critical thermodynamic region, where the temperature remains constant. For the pressure $P>P_{c}$, as the radius of the thermal profile increases, the BH temperature decreases. Moreover, Fig.~\ref{fig5} shows that for a fixed magnetic charge $g$, the temperature increases with the pressure while the region of the thermal profile decreases with the pressure. In order to study the N-type change trend of the Hayward-AdS BH with different magnetic charges, the thermal profiles corresponding to the region from $r_{s1}$ to $r_{s2}$ are plotted in Fig.~\ref{fig6}. The results show that as the radius of the thermal profile increases, the temperature increases at the beginning, then decreases, and increases again at the end. Hence, the phase structure of the Hayward-AdS BH can be reflected by the thermal profile when the pressure is smaller than the critical pressure. 

\vspace{10pt}

\noindent {\bf Acknowledgments}

\noindent
This work was supported by the National Natural Science Foundation of China (Grant Nos. 11873001 and 12047564), the Fundamental Research Funds for the Central Universities (Grand No. 2021CDJZYJH-003), and the Postdoctoral Science Foundation of Chongqing (Grant No. cstc2021jcyj-bsh0124).

\end{document}